\documentclass{aastex6}
\usepackage{amsmath}
\usepackage{amsfonts}
\usepackage{amssymb}
\usepackage{multirow}
\usepackage{graphicx}
\usepackage{booktabs}

\newcommand{\adsurl}[1]{\href{#1}{ADS}} 
\providecommand{\url}[1]{\href{#1}{#1}}

\def \apjs {{\rm {ApJS}}}

\def \aap {{\rm {A\&A}}}

\def \apjl{\rm {ApJL}}

\def \deg{$^\circ$}

\begin{document}
\title{Color difference makes a difference: four planet candidates around $\tau$ Ceti}
\author{F. Feng\altaffilmark{1}, M. Tuomi, H. R. A. Jones}
\affil{Centre for Astrophysics Research, School of Physics, Astronomy and Mathematics, University of Hertfordshire, College Lane, Hatfield AL10 9AB, UK}
\author{J. Barnes}
\affil{Department of Physical Sciences, The Open University, Walton Hall, Milton Keynes MK7 6AA, UK}
\author{G. Anglada-Escud\'e}
\affil{School of Physics and Astronomy, Queen Mary University of London, 327 Mile End Road, E1 4NS, London, UK}
\author{S. S. Vogt}
\affil{UCO/Lick Observatory, Department of Astronomy and Astrophysics, University of California at Santa Cruz, CA 95064, USA}
\and 
\author{R. P. Butler}
\affil{Department of Terrestrial Magnetism, Carnegie Institute of Washington, Washington, DC 20015, USA}

\date{\today}

\altaffiltext{1}{fengfabo@gmail or f.feng@herts.ac.uk}

\begin{abstract}
  The removal of noise typically correlated in time and wavelength is one of the main challenges for using the radial velocity method to detect Earth analogues. We analyze radial velocity data of $\tau$ Ceti and find robust evidence for wavelength dependent noise. We find this noise can be modeled by a combination of moving average models and ``differential radial velocities''. We apply this noise model to various radial velocity data sets for $\tau$ Ceti, and find four periodic signals at 20.0, 49.3, 160 and 642\,d which we interpret as planets. We identify two new signals with orbital periods of 20.0 and 49.3\,d while the other two previously suspected signals around 160 and 600\,d are quantified to a higher precision. The 20.0\,d candidate is independently detected in KECK data. All planets detected in this work have minimum masses less than 4$M_\oplus$ with the two long period ones located around the inner and outer edges of the habitable zone, respectively. We find that the instrumental noise gives rise to a precision limit of the HARPS around 0.2 m/s. We also find correlation between the HARPS data and the central moments of the spectral line profile at around 0.5\,m/s level, although these central moments may contain both noise and signals. The signals detected in this work have semi-amplitudes as low as 0.3\,m/s, demonstrating the ability of the radial velocity technique to detect relatively weak signals. 
\end{abstract} 
\section{Introduction}     \label{sec:introduction}
The radial velocity (RV) technique is one of the most successful methods used to detect exoplanets. The extreme precision spectrographs developed in recent years have improved the precision of Doppler measurements down to a few meters per second. In particular, the High Accuracy Radial Velocity Planet Searcher (HARPS) spectrometer has enabled the discovery of Super-Earths due to its precision of measuring down to 1\,m/s RV \citep{pepe02,mayor03}. However, this precision is still not
high enough to detect Earth analogs in the habitable zone of nearby stars, which
requires achieving 10\,cm/s precision \citep{mayor14}. Moreover, efficient statistical tools and noise models are required to entangle the signals from stellar and instrumental noise, as summarized in the results of RV challenge \citep{dumusque16b}. 

The Keplerian signals in RV measurements can be diluted
and distorted by stellar activity, rotation and uneven sampling of
observation times. These sources of contamination can be partly removed by various activity indicators such as the Ca II HK emission (RHK), line bisector span (BIS) and the width of the spectral lines (FWHM). However, the relation between the indicators and their RV counterparts could be very complex and is not necessarily deterministic, leading to controversial results in the validation of planetary candidates (e.g., \citealt{robertson14,anglada15}). This incomplete modeling of
RV noise together with a lack of consensus on the most appropriate and efficient statistical methods are limiting the abilities of RV analysis to detect Earth analogues (see \citealt{feng16}  for details). Nevertheless, our noise modeling approach is one of the best RV modeling strategies in the field according to the analysis of the results of RV fitting challenge \citep{dumusque16b}.

Another challenge is the dependence of RV noise on wavelength or, in practice, on echelle order because RVs for HARPS are determined on an order-by-order basis. Since the jitter in RV variations depends on spectral orders \citep{anglada12}, the RV averaged over all orders would contain wavelength dependent noise due to a lack of appropriate weighting and correcting. This motivates us to model the color dependency of the RVs. We divide the 72 spectral orders into groups, and average the RVs in each group to generate the so-called ``aperture data sets'' and investigate the differences between these aperture data sets -- the so-called ``differential RVs'' \citep{feng17a}.

In this work, we use a combination of moving average models and the differential RVs to remove wavelength and time dependent noise. We apply this model to the HARPS measurements of $\tau$ Ceti, which may host a multi-planetary system according to previous analyses (\citealt{tuomi12}; hereafter MT13). $\tau$ Ceti is a Sun-like star but is not as active as the Sun. There are currently more than 9000 HARPS measurements of this star, potentially enabling us to find signals with semi-amplitude as low as 0.2\,m/s (MT13). Although MT13 have removed part of the correlated noise using moving average models, their noise modeling is probably incomplete, since the wavelength dependent noise was not taken into account. With new data obtained by HARPS and the use of differential RVs, we reanalyze the RV variations of $\tau$ Ceti to find Keplerian signals and to attempt to verify the results of MT13. 

This paper is structured as follows. First, we introduce the HARPS and KECK measurements of $\tau$ Ceti, and define various data sets in section \ref{sec:data}. In section \ref{sec:method} we describe the Markov Chain Monte Carlo (MCMC) method used to sample the posterior distribution within the Bayesian framework. Then we justify the use of differential RVs to remove wavelength dependent noise, and select the optimal noise model for each data set in section \ref{sec:dv}. We apply these models to find planetary candidates, and compare them with previous results in section \ref{sec:signals}. We also investigate the cause of highly eccentric signals. In section \ref{sec:candidates}, we report the parameters of planetary candidates, and analyze the dynamical stability and habitability of these planetary candidates. Finally, we discuss and conclude in section \ref{sec:conclusion}. 
\section{Radial velocity Data of $\tau$ Ceti}\label{sec:data}

In the European Southern Observatory archive, there are more than 9000 publicly available RVs measured by HARPS from June 2003 to September 2013 for HD10700 as part of the observing programs 60.A-9036, Mayor, Comm, 072.C-0488, 072.C-0513, 074.D-0380, 075.C-0234, 075.D-0760, 076.C-0073, 077.C-0530, 078.C-0751, 078.C-0833, 079.C-0681, 081.C-0034, 082.C-0315, 083.C-1001, 084.C-0229, 085.C-0318, 086.C-0230, 087.C-0990, 088.C-0011, 089.C-0050, 090.C-0849 and  091.C-0936.

The main data we will use are the RVs measured by the HARPS \citep{mayor03} and processed by the TERRA pipeline \citep{anglada12}. The data are processed using the astrocatalog mode of TERRA whereby all barycentric corrections are recomputed using consistent ephemeris and coordinates and proper motions based on \cite{leeuwen07}. This means that the calculation of barycentric earth radial velocity does not rely on telescope header information input by the different HARPS programmes that we have used data from. The TERRA algorithm also produces 72 data sets, one for each HARPS spectral order. Each of them is composed of RVs measured at a certain wavelength range. The RVs are analyzed in combination with three activity indices, including the calcium activity index (S-index or $I_S$), the spectral line bisector (BIS or $I_B$) and full width at half maximum (FWHM or $I_F$) of the cross-correlation function (CCF). To increase the signal to noise ratio in aperture data sets, we evenly divide the 72 RV orders into groups, and average the data sets by order in each group weighted by their measurement uncertainties to form an averaged data set. For example, we divide the 72 orders into $n$ groups, and average the data sets in each group to generate $n$ aperture data sets, named $n$AP$i$, where $i=1,...,n$. The average of all orders forms the 1AP1 data set. 

To remove short term noise in the RV data sets, we define another type of RVs by binning the RVs measured within one hour. We start from the beginning of an RV data set and set the beginning time as the reference time. Then we average the RVs within one hour from the reference time weighted by their uncertainties. We then define the first time point out of the one hour window as the next reference time, and average the RVs within the one hour window in the same way. Repeating this, we generate the binned version of a given RV data set. The binned version of the $n$AP$i$ data set is dubbed ``binned$n$AP$i$''. 

The outliers beyond 5-$\sigma$ of the RVs in 1AP1 are removed from all aperture data sets. Considering that the noise caused by stellar activity may not be properly estimated by measurement errors (including photon poisson noise and a calibration error of 30\,cm/s), we also weight each data set by a constant based on the sum of jitter and measurement uncertainty terms. We try different jitter levels and do not find significant changes in the periodograms of the aperture data sets. In other words, the signals in the aperture data sets are not sensitive to the weighting function used to average spectral orders. Thus we still use the measurement errors to weight data sets in the averaging process.

We define the RV differences between aperture data sets as differential RVs. We denote them by ``nAPx-x$'$'', where n is the number of divisions, and x and x$'$ denote different data sets in the n divisions of the 72 aperture data sets. For example, by subtracting 3AP1 from 3AP2, we obtain the 3AP2-1 data set. We will use differential RVs to remove the wavelength dependent noise in section \ref{sec:dv}. 

\begin{figure}
  \centering
  \includegraphics[scale=0.7]{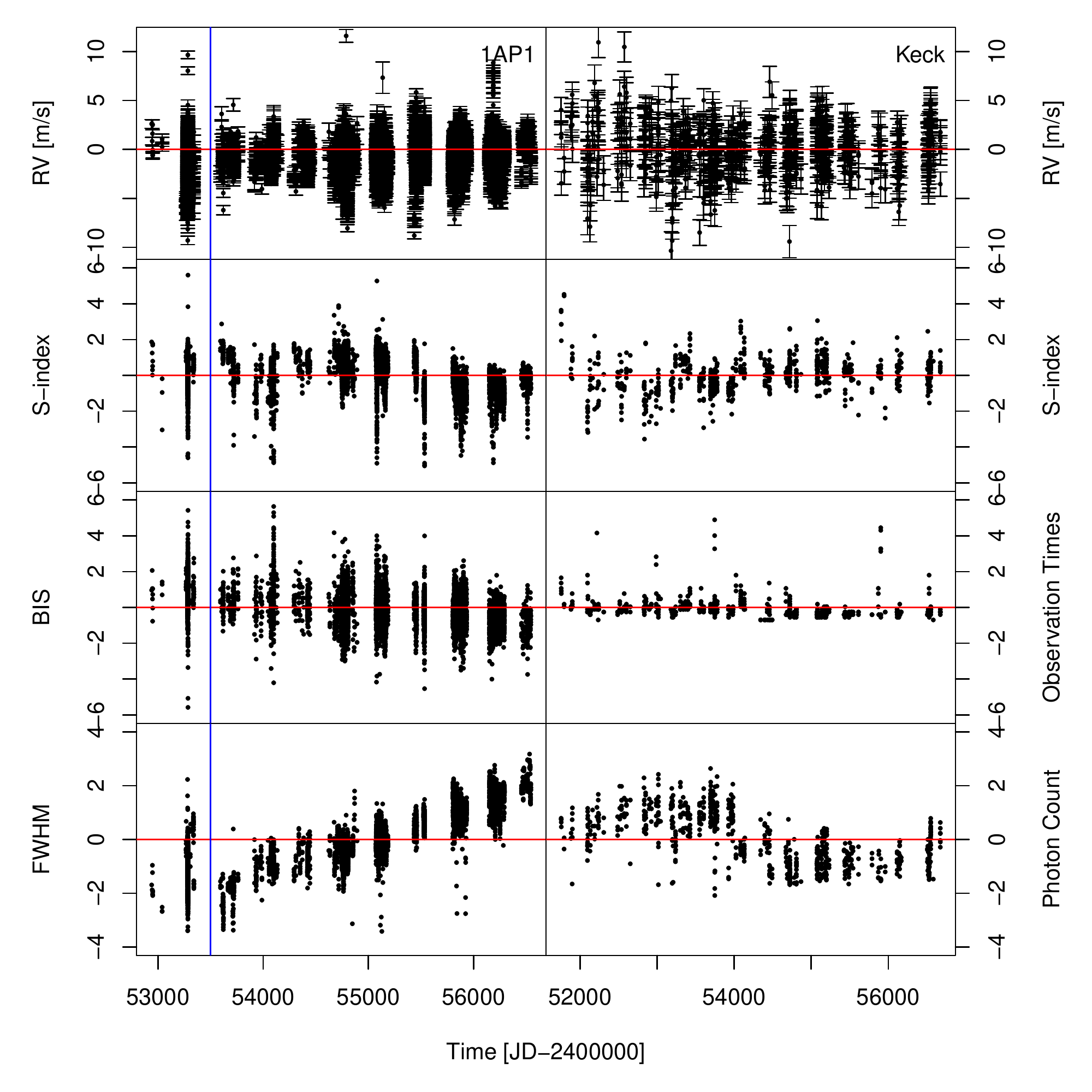}
  \caption{The RV data and activity indicators of 1AP1 or HARPS (left panels) and KECK (right panels). The red lines denote the zero RV, and the blue line shows the epoch of JD2453500 which is used to define the C1AP1 data set. The activity indices are normalized to the zero mean and unit standard deviation. The measurement uncertainties of the RV data are shown with error bars. There are 8880 and 752 data points in the 1AP1 and KECK data sets, respectively.}
  \label{fig:data}
\end{figure}

\begin{figure}
  \centering
  \includegraphics[scale=0.2]{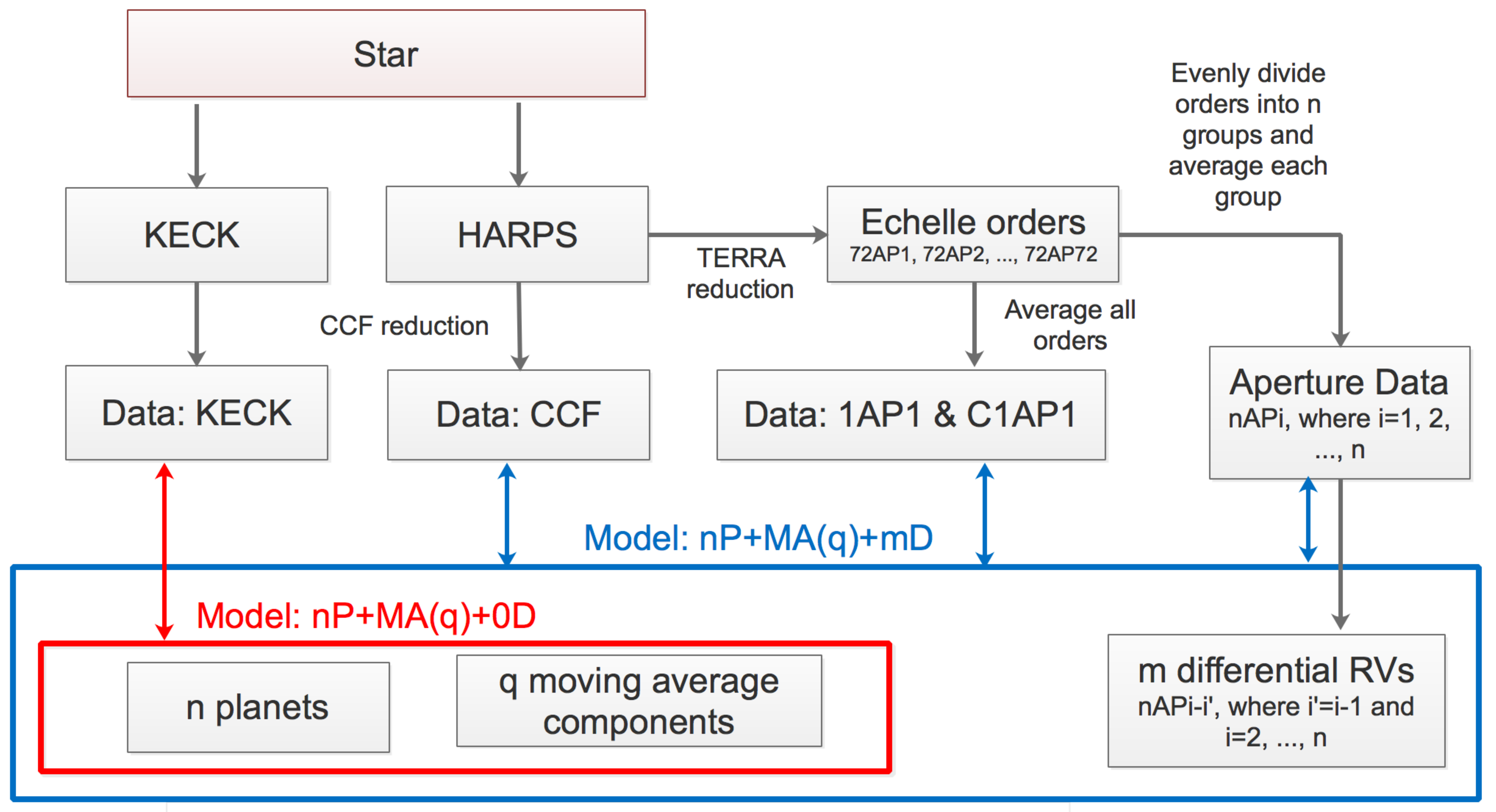}
  \caption{The data reduction and modeling process.}
  \label{fig:diagram}
\end{figure}

Apart from the TERRA-reduced HARPS measurements of $\tau$ Ceti, we also use the HARPS data reduced by the CCF method and the RVs measured by the HIRES spectrometer on the KECK telescope \citep{butler16}. For HIRES we model the dependence of RV variation on the photon count and integration time. The 1AP1 and KECK data sets and their normalized activity indices are shown in Fig. \ref{fig:data}. These data together with the aperture data sets are published electronically. Around JD2453280, the FWHM scatter greatly and the RV changes rapidly. From JD2453280 to JD2453285, $\tau$ Ceti was observed for asteroseismology purpose. The star was observed continuously for 5 days, with exposure times of 40\,s \citep{teixeira09}. For such short exposure time and high-cadence measurements, the data is contaminated by excess noise from a periodic guiding error \citep{teixeira09}. So we remove the 1597 data points before JD2453500 to form a more conservative subset named ``C1AP1''. The whole process of data reduction and modeling is shown in Fig. \ref{fig:diagram}. 

\section{Data analysis methods}\label{sec:method}

We compare RV models in the Bayesian framework. We start from the Bayes theorem which is
\begin{equation}
  P(M_i|\mathcal{D})= \frac{P(\mathcal{D}|M_i)P(M_i)}{\sum\limits_{j} P(\mathcal{D}|M_j)}~, 
\label{eqn:BF}
\end{equation}
where $M_i$ is a model, and $\mathcal{D}$ is the data, $P(M_i|\mathcal{D})$ is the posterior, $P(\mathcal{D}|M_i)$ and $P(M_i)$ are the evidence (or marginalized likelihood) and the prior of model $M_i$, respectively. The denominator of Eqn. \ref{eqn:BF} is a normalization term. The posterior of a model is a measure of its plausibility. If no model is preferred {\it a priori}, the posterior ratio is equal to the evidence ratio, which is also called ``Bayes factor'' (BF). We claim that a model is favored over another if the BF is larger than a certain value. According to the analyses of the RVs for M dwarfs by \cite{feng16}, the BF estimated by the Bayesian information criterion (BIC) combined with a threshold of 150 avoids false positives and negatives\footnote{Although the comparison of BF estimators performed by \cite{feng16} is for M dwarfs, the BIC is found to be rather conservative, and thus is appropriate for $\tau$ Ceti which is a quiet star.}. This is equivalent to the criterion, $\Delta{\rm BIC}>10$, suggested by \cite{kass95}, although they did not test this for RV data we have found it too work well. Thus we use the BIC to estimate the BF, and select signals using the BF threshold of 150 in this work.  

For a given model M and data $\mathcal{D}$, we need to estimate its parameters $\rm{\theta}$ by calculating their posterior densities according to 
\begin{equation}
  P({\bf\theta}|\mathcal{D},M)  =  \frac{P(\mathcal{D}|{\bf\theta},M) P({\bf \theta}|M)}{P(\mathcal{D}|M)}~,
\label{eqn:posterior}
\end{equation}
where $P(\mathcal{D}|{\bf\theta},M)$ and $P({\bf\theta}|M)$ are the likelihood and prior distributions, respectively. The specific likelihood and prior distributions for various models will be introduced in the next section. Because the posterior distributions for RV models are always multi-dimensional and multi-modal, the prior sampling may not well resolve the narrow posterior maxima. Thus we sample the posterior using the MCMC implemented by the adaptive Metropolis-Hastings algorithm \citep{haario01}, which was first applied to the analysis of RV data by \cite{tuomi12c}. We first launch tempered/hot chains to explore the whole parameter space and to find local posterior maxima. We then generate untempered/cold chains to explore these maxima in order to quantify the signals and to estimate the parameters ``maximum {\it a posteriori}'' (MAP). Our method is similar to the parallel tempering MCMC algorithm introduced by \cite{gregory05}. 

To make the MCMC chains identify the most probable areas of the posterior, we generate hot and cold chains in the following way. First, we run cold chains to obtain the posterior density for the null hypothesis that no signals exist in the data. Second, we evenly divide the logarithm of the period range into 20 intervals, and launch 4 hot chains to find posterior maximum for each. We call all of the maxima found by these chains as primary signals. Third, we choose the primary signal that gives the highest likelihood, and start cold chains from the position of this signal to generate a statistically representative posterior sample. Fourth, we compare the one-planet model with the null hypothesis using the BF threshold. If the one-planet model is favored, we move on to add another primary signal into the model, and run MCMC chains to calculate the BF for model comparison. We run steps 3-4 repeatedly until an extra Keplerian signal is not favored by the data. In addition to the BF threshold, we follow \cite{tuomi12b} to confirm a signal if it is constrained from above and below in the marginalized posterior distribution , i.e. $P(P_k|\mathcal{D},~M)$ converges to a stationary distribution.

\section{Differential radial velocities}\label{sec:dv}
\subsection{Wavelength dependent noise models}\label{sec:model}
The jitter in RV variations is probably wavelength dependent due to the wavelength dependent modulation of stellar radiation by stellar activity \citep{huelamo08, desort07}. This dependence indicates that the RVs determined through averaging RV orders are biased due to a lack of proper weighting and correction. To avoid this bias and model wavelength dependent RVs, we introduce a new
type of noise proxy called ``differential RVs'', defined by the RV differences between aperture data sets. Since Keplerian signals do not depend on wavelength, the differential RVs only contain wavelength dependent noise. Actually, they are approximations of the derivatives of RV noise with respect to wavelength. They provide important information about stellar activity and instrumental noise, and thus can be used to remove the wavelength dependent noise in Doppler measurements. 

If we express the non-Keplerian part in an aperture data set with wavelength range centered at $\lambda$ as $\Psi(t,\lambda)$, the differential RV is defined by the difference between two aperture data sets,
\begin{equation}
  D(t_i,\lambda_j)\equiv v(t_i,\lambda_{j+1})-v(t_i,\lambda_j)=\Psi(t_i,\lambda_{j+1})-\Psi(t_i,\lambda_j)~,
\end{equation}
where $v(t_i,\lambda_j)$ is the RV at time $t_i$ and wavelength $\lambda_j$. 
If the wavelength difference is small enough, $D(t_i,\lambda_j)$ is approximately proportional to the first partial derivative of $\Psi$ with respect to $\lambda_j$,
\begin{equation}
\left. {\frac{\partial \Psi}{\partial \lambda}}\right|_{\lambda=\lambda_j}\approx\frac{\Psi(t_i,\lambda_{j+1})-\Psi(t_i,\lambda_j)}{\lambda_{j+1}-\lambda_j}\propto
  D(t_i,\lambda_j)~. 
  \label{eqn:1order}
\end{equation}
Similarly, if the wavelength difference is a constant ($\Delta\lambda$) for
all differential RVs, the second partial derivative of $\Psi$ is approximately 
\begin{equation}
  \left.{\frac{\partial^2 \Psi}{\partial
    \lambda^2}}\right|_{\lambda=\lambda_j}\approx\frac{\Psi(t_i,\lambda_{j+1})+\Psi(t_i,\lambda_{j-1})-2\Psi(t_i,\lambda_j)}{\Delta\lambda^2}\propto
  D(t_i,\lambda_j)-D(t_i,\lambda_{j-1})~. 
  \label{eqn:2order}
\end{equation}
In principle, we can continue the above calculations to derive higher order
derivatives of $\Psi$ from differential RVs. These derivatives can be
used to approximately reconstruct the wavelength dependent noise at a given
time. For example, if the dependence of noise on time and wavelength is described by a quadratic polynomial function with time-varying parameters, i.e.
\begin{equation}
  \Psi(t,\lambda)=p(t)\lambda^2+q(t)\lambda+r(t)~,
\end{equation}
the differential RVs could approximate $p(t)$ and $q(t)$ according to
\ref{eqn:1order} and \ref{eqn:2order}. Hence the discrete form of
the above equation is
\begin{equation}
  \Psi(t_i,\lambda_j)\approx
  \frac{[D(t_i,\lambda_{j+1})-D(t_i,\lambda_{j-1})]\lambda_j^2}{2\Delta\lambda^2}+\frac{D(t_i,\lambda_j)\lambda_j}{\Delta\lambda}+r(t_i)~.
  \label{eqn:poly_psi}
\end{equation}

Considering that the dependence of noise on time and wavelength is probably much more complex than the above case, we estimate the noise using the
following equation, 
\begin{equation}
\hat{\Psi}(t_i,\lambda_j) =
a(\lambda_j)\,t_i+b(\lambda_j)+\sum_{k}c_k(\lambda_j)I_k+\sum_{m=1}^{N_\lambda-1}d_mD(t_i,\lambda_m)~,
\label{eqn:psi}
\end{equation}
where $N_\lambda$ is the number of aperture divisions and thus $N_\lambda-1$ is
the number of independent differential RVs, $d_m$ characterizes the linear dependence of RV noise on differential RV at wavelength $\lambda_m$ which is averaged over the differential RV, $a$ is the acceleration caused by activity cycles or other wavelength dependent noise \footnote{For example, the atmospheric absorption and/or scattering of the star light could be wavelength dependent, leading to wavelength dependent noise in RVs. Considering that the linear trend in the model may be caused by stellar activity and instrumental noise, we put it in the noise component of the RV model.} and $b$ is the reference velocity. The linear dependence of RVs on activity index $I_k$ is parameterized by constant $c_k$. All these parameters are wavelength dependent up to a certain level. This equation is equivalent to a set of independent noise models applied to different aperture data sets. 

To predict the RV variation, we combine $n$ Keplerian components with a wavelength dependent noise  component to form a basic RV model which is 
\begin{eqnarray}
  \hat{v}_b(t_i, \lambda_j)&=&\sum_{k=1}^{n} f_k(t_i)+\hat{\Psi}( t_i,\lambda_j)~,\\\nonumber
  f_k(t_i)&=&K_k [\cos(\omega_k + \nu_k(t_i))+e_k\cos(\omega_k)]~,
\label{eqn:basic}
\end{eqnarray}
where $f_k(t_i)$ is the RV variation caused by the $k^{\mathrm{th}}$
planet, and $\hat{\Psi}$ is the estimation of the noise component. In the Keplerian component, $K_k$, $\omega_k$, $\nu_k$, $e_k$  are the
amplitude, longitude of periastron, true anomaly and eccentricity for the
$k^{\mathrm{th}}$ planetary signal. 

According to \cite{feng16}, the time-varying RV noise ($r(t)$ in our
case) of M dwarfs can be well modeled using a combination of white noise and the first order moving average (MA) models. However, $\tau$ Ceti is much hotter than M dwarfs and thus the convective velocity is greater, potentially giving rise to significant granulation and stellar oscillation signals (e.g., \citealt{meunier17, kjeldsen95}). Moreover, the data is sampled with high cadence and thus probably contaminated by significant correlated noise. Thus we consider higher order moving average models to remove the RV noise, following MT13. We define the general moving average model with exponential smoothing as
\begin{equation}
  \hat{v}(t_i,\lambda_j)=\hat{v}_b(t_i,\lambda_j)+\sum_k w_k(\lambda_j) \exp[-|t_i-t_{i-k}|/\tau(\lambda_j)] \epsilon_{i-k},
\label{eqn:full}
\end{equation}
where $w_k$ and $\tau$ are the amplitude and time scale of the moving
average while $\epsilon_{i-k}$ is the residual after subtracting the data by
a realization of the basic model at time $t_{i-k}$. Hereafter, we define ``nP+MA(q)+mD'' as the $n$-planet model with $q^{\rm{th}}$-order moving average and $m$ differential RVs which are derived from $m+1$ aperture data sets (see Fig. \ref{fig:diagram}). The white noise model is denoted by MA(0). 

For the 1AP1 data sets, the wavelength $\lambda_j$ should be regarded as an averaged wavelength. Eqn. \ref{eqn:basic} is still applicable because the averaging process is a linear process, and thus the linearity of the RV model remains. Although the moving average model has an exponential term, the correlation time scale $\tau$ is not sensitive to wavelength according to our analysis. Hence we will apply Eqn. \ref{eqn:basic} to 1AP1 and other aperture data sets.

For a given aperture data set, the excess white noise or white jitter is taken into account in the likelihood, 
\begin{equation}
\mathcal{L}_j\equiv P(v_j|\boldsymbol{\theta},M)=\prod_i\frac{1}{\sqrt{2\pi[\sigma_{i,j}^2+s_J(\lambda_j)^2]}}\exp\left[-\frac{(\hat{v} (t_i,\lambda_j)-v_{i,j})^2}{2(\sigma_{i,j}^2+s_J(\lambda_j)^2)}\right]~,
\label{eqn:like}
\end{equation}
where $\sigma_{i,j}$ is the measurement noise at time $t_i$ in the j$^{th}$ aperture data set which has an averaged wavelength of $\lambda_j$, $s_J(\lambda_j)$ is
the jitter level, and $v_{i,j}$ is the observed RV at time $t_i$ in the j$^{th}$ aperture data set.

Following \cite{feng16}, we adopt uniform prior distributions for most 
parameters except for the eccentricity and some time scale 
parameters. Since the planets with highly eccentric orbits are very 
rare, we adopt a Gaussian distribution centered at zero and with a 
standard deviation of 0.1 \citep{tuomi13}. Although we set a lower limit of one day for the orbital period, we do investigate shorter period signals if the power of short period is high in the periodogram. The prior distributions of all parameters are described in Table \ref{tab:prior}. 

\begin{table}
  \centering
\caption{The prior distributions of model parameters. The 
  unit of $c_k$ and $d_m$ is m/s because the activity indices and differential RVs are normalized to zero mean and unit standard deviation before inclusion in the model. The maximum and minimum time of the RV data are denoted by $t_{\textrm{max}}$ and $t_{\textrm{min}}$, respectively. The maximum amplitude of the RV data set with respect to the mean is denoted as $|v-\bar{v}|_{\rm{max}}$. The parameter characterizing the dependence of RV on activity indices is $c_k$, where $k$ denotes the names of various indices. }
\label{tab:prior}
  \begin{tabular}  {c*{4}{c}}
\hline 
Parameter&Unit&Prior distribution& Minimum & Maximum\\\hline 
        \multicolumn{5}{c}{\it Each Keplerian signal}\\
$K_j$&m/s&$1/(K_{\text{max}}-K_{\text{min}})$&0&$2|v-\bar{v}|_{\text{max}}$\\
$P_j$&day&$P_j^{-1}/\log(P_{\text{max}}/P_{\text{min}})$&1&$t_{\text{max}}-t_{\text{min}}$\\
$e_j$&---&$\mathcal{N}(0,0.1)$&0&1\\
$\omega_j$&rad&$1/(2\pi)$&0&$2\pi$\\
$M_{0j}$&rad&$1/(2\pi)$&0&$2\pi$\\\hline 
        \multicolumn{5}{c}{\it Linear trend and jitter}\\
$a$&m\,s$^{-1}$yr$^{-1}$&$1/( a_{\text{max}}- a_{\text{min}})$&$-365.24K_{\text{max}}/P_{\text{max}}$&$365.24K_{\text{max}}/P_{\text{max}}$\\
    $b$&m/s&$1/( b_{\text{max}}- b_{\text{min}})$&$-K_{\text{max}}$&$K_{\text{max}}$\\
$s_J$ &m/s&$1/(s_{J\rm max}-s_{J\rm min})$&0&$K_{\rm max}$\\\hline 
\multicolumn{5}{c}{\it Moving average}\\ 
$w$&---&$1/(w_{\text{max}}-w_{\text{min}})$&-1&1\\
$\tau$&day&$P_j^{-1}/\log(\tau_{\text{max}}/\tau_{\rm{min}})$&$1/(t_{\rm{max}}-t_{\rm{min}})$&1\\\hline 
    \multicolumn{5}{c}{\it Activity indices and differential RVs}\\
$c_k$&m/s&$1/(c_{k\text{max}}-c_{k\text{min}})$&$-c_{k\text{max}}$&$K_{\text{max}}/(I_{X\text{max}}-I_{X\text{min}})$\\
$d_m$\,($m\in\{1,...,N_\lambda-1\}$)&m/s&$1/(d_{m\rm{max}}-d_{m\text{min}})$&$-d_{m\rm{max}}$&$K_{\text{max}}/(D_{m\rm{max}}-D_{m\rm{min}})$\\\hline 
  \end{tabular}  
\end{table}

Nearly all radial velocity data analysis has only considered the averaged data or 1AP1 is analyzed without accounting for wavelength dependent noise. Here, we demonstrate the wavelength dependence of RV noise by applying the model defined by Eqn. \ref{eqn:full} to aperture data sets in subsection \ref{sec:remove} and \ref{sec:instrumental}. In the other sections, we will apply Eqn. \ref{eqn:full} to the 1AP1 and other data sets in order to identify Keplerian signals. The number of Keplerian components, MA components and differential RVs will be chosen based on Bayesian model comparison. 

\subsection{Removing wavelength dependent noise}\label{sec:remove}

To see whether the differential RVs can reduce the wavelength dependence of RV noise, we compare RV models with and without differential RVs in the Bayesian framework. We apply the RV model defined in Eqn. \ref{eqn:full} and \ref{eqn:like} to the 3AP1, 3AP2, 3AP3 and 1AP1 data sets. Specifically, we model the aperture data sets using the first order moving average and white noise models without and with dependence on differential RVs, which are denoted by 0P+MA(1)+0D, 0P+ MA(1)+2D, 0P+MA(0)+0D, 0P+MA(0)+2D, respectively. To visualize the differences between various noise models, we calculate the generalized periodograms for the data sets and their residuals.

The generalized periodogram is calculated by maximizing the logarithmic likelihood of a combination of sinusoidal functions and a linear trend at a sample of frequencies. Unlike the Lomb-Scargle periodogram, this general periodogram not only optimizes the amplitude and phase in the sinusoidal function but also optimizes the reference velocity and linear acceleration for each frequency. We call this periodogram the generalized Lomb-Scargle periodogram with floating trend (GLST\footnote{The code for calculating GLST is available at \url{https://github.com/phillippro/agatha} and a corresponding online app is at \url{http://www.agatha.herts.ac.uk}.}; \citealt{feng17b}; also see \citealt{baluev08} and \citealt{suveges15}), a generalization of the so-called generalized Lomb-Scargle periodogram (GLS; \citealt{zechmeister09}).  Following \cite{cumming99}, the GLST is normalized using the residual variance, and the FAP values are calculated accordingly. They FAP is not accurate for signal identification/quantification especially when the RV noise is highly correlated in time and wavelength. Thus we only show FAPs in the GLSTs of residuals where correlated noise is properly modeled and subtracted. Although there are calculations of more precise FAPs \citep{baluev08} and more sophisticated periodograms like the ones introduced by \cite{baluev13b,baluev15} and by \cite{feng17b}, we rely on Bayesian posterior samplings to identify and quantify signals. 

\begin{figure}
  \centering
  \hspace*{-10mm}
\includegraphics[scale=0.3]{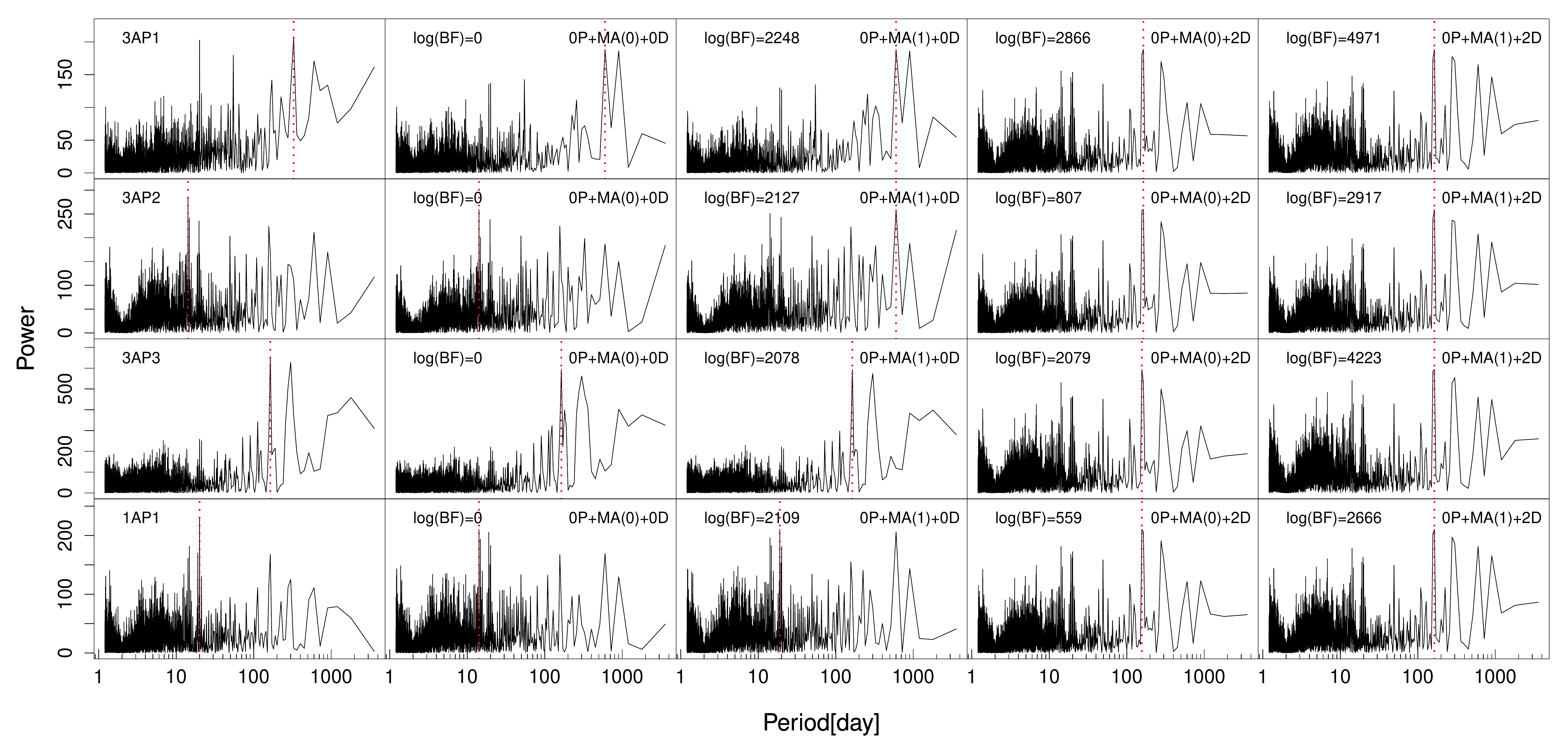}
  \caption{We aim to illustrate the existence of wavelength-dependent noise and the necessity of removing it using differential RVs. The interpretation of specific peaks in plotted periodograms is not important. Rather, our concern is the consistency between periodograms of RVs and their residuals measured at different wavelengths. Thus we show a series of different periodograms across the page. Each row of plots is given to a different wavelength range. Going down the page they are for the datasets 3AP1, 3AP2, 3AP3 (splitting the data set into three parts) and 1AP1 (regular data set averaged over all orders). The following columns show periodograms for the corresponding residuals after subtracting the model predictions for the 0P+MA(0)+0D (white noise), 0P+MA(1)+0D (moving average), 0P+MA(0)+2D (white noise with differential RVs), 0P+MA(1)+2D (moving average with differential RVs). The logarithm BF of a model with respect to the MA(0) model are shown for the residuals of each data set after subtracting the best model prediction. The red dotted lines denotes the periods at the maxima of posterior distributions.}
\label{fig:residual}
\end{figure}

The periodograms for the 3AP1, 3AP2, 3AP3 and 1AP1 data sets and their residuals after subtracting the model prediction are shown in Fig. \ref{fig:residual}.
Notably we observe great differences between periodograms for the data sets of 3AP1, 3AP2, 3AP3 and 1AP1 which indicates that wavelength dependent noise is an important factor. On the contrary, the residuals of all RV data sets have similar periodograms after subtracting the best predictions of 0P+MA(0)+2D or 0P+MA(1)+2D from the data, indicating the essential role of differential RVs in removing wavelength dependent noise. In the right two columns of panels in Fig. \ref{fig:residual}, we see consistent periodograms for residuals despite being calculated for different noise models and aperture data sets. 

For the 1AP1 data set, the differential RVs do not improve the BF as much as the moving average model. On the other hand, the increase of logarithmic BF by including both MA(1) and differential RVs in the model is approximately equal to the sum of those by including them separately (see the left three columns). This indicates that the wavelength dependent noise and time-correlated noise are independent and thus should be modeled with independent noise components. The inclusion of differential RVs improves the fitting more for the 3AP1 and 3AP3 data sets than it does for the 3AP2 and 1AP1 data sets. This means that the RVs measured at the middle of the wavelength range contain less wavelength-dependent noise than those measured at the blue and red ends of the range do. The 1AP1 data set contains less correlated noise probably due to a partial removal of wavelength-dependent noise by the averaging of all spectral orders. But this noise is still significant enough to contaminate the periodogram and thus result in detections of noise-induced signals.

In summary, for both the 1AP1 and the other data sets, the differential RVs are able to remove wavelength dependent noise and help avoiding false positives. This role of differential RVs is rather different to that of red noise models, which are good at removing time-correlated but not wavelength-correlated noise. We also conduct similar analysis for the binned data sets, and find similar results, although the wavelength dependent noise is greatly reduced by the binning process. Considering that $\tau$ Ceti is a quiet main sequence star, our analysis is probably representative. The wavelength dependence of RV noise is probably stronger for more active stars such as Alpha Centauri A/B \citep{dumusque12}. For them, differential RVs would play a key role in detecting exoplanets consistently. 

\begin{figure}\centering 
\includegraphics[scale=0.6]{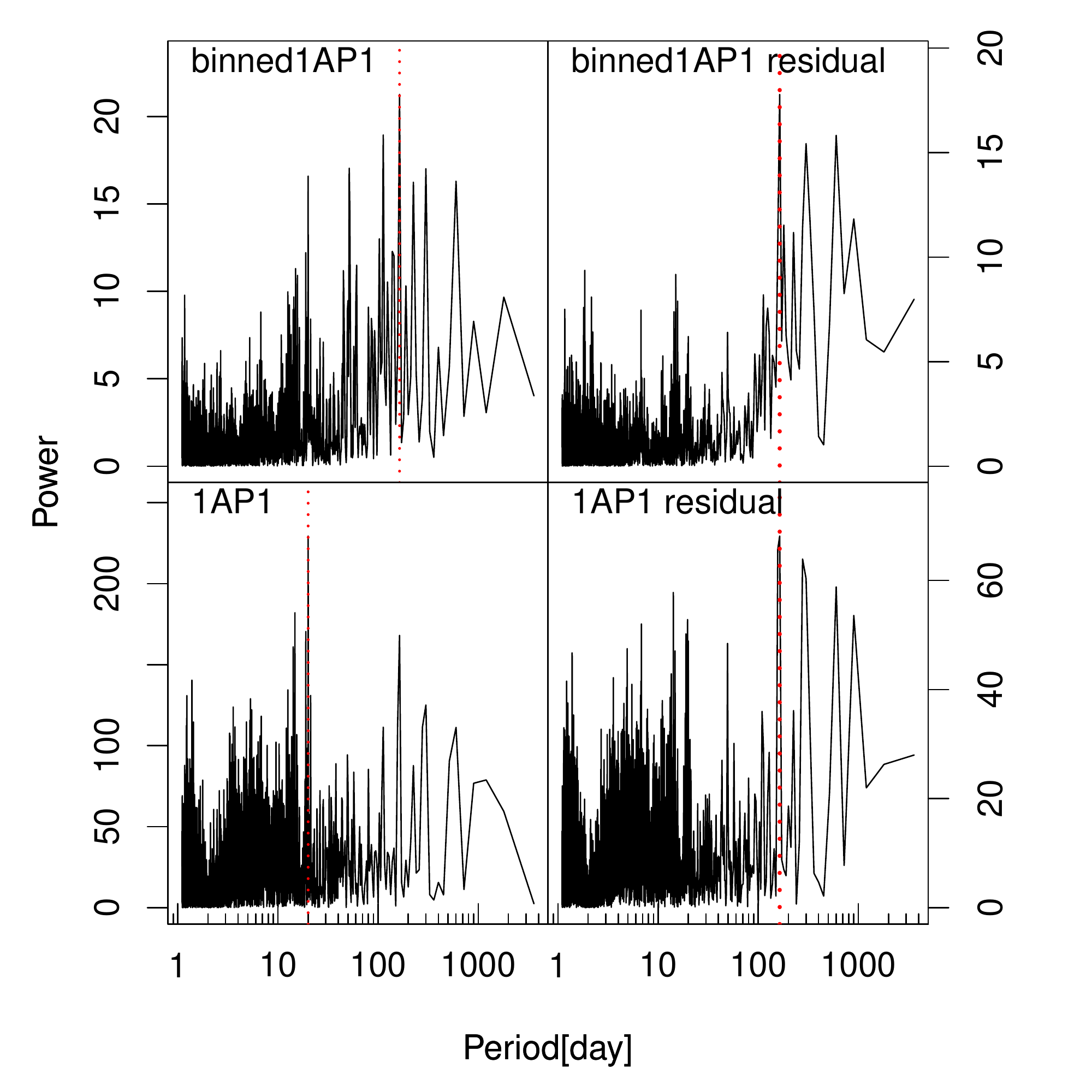}
  \caption{Similar to Fig. \ref{fig:residual}, but for the binned1AP1 (top left) and 1AP1 (bottom left) data sets and for their residuals (right panels) calculated by subtracting the 0P+MA(1)+2D prediction from the data. }
\label{fig:binning_effect}
\end{figure}

In the following sections, we will focus our analysis on the 1AP1 data set because it has higher signal to noise than aperture data sets, and has residuals similar to those of aperture data sets. We further study the effect of binning by comparing the periodograms for the binned1AP1 and 1AP1 data sets and for their residuals in Fig. \ref{fig:binning_effect}. We find that periodograms are very different both for the RV data sets and for their residuals. The binning of data over one hour time span has removed features caused by noise and signals altogether. This makes the binned version unreliable for detecting signals. 

\subsection{Modeling instrumental noise}\label{sec:instrumental}
Like stellar activity, instrumental noise may also be colorful because of the potential nonlinear response of instrument to wavelength. To model  the instrumental noise in the data, we generate the calibration data sets by combining the RV aperture data sets measured by the HARPS and reduced by the TERRA algorithm for 172 stars which have been reduced with the TERRA reduction and representing the most frequently observed targets (excluding $\tau$ Ceti). Most of these RVs were measured within the HARPS-Upgrade GTO program \citep{pepe11,mayor11}. We derive aperture data sets and differential RVs from these HARPS measurements, and combine them. Specifically, we remove the (differential) RVs which have absolute values larger than 20\,m/s or deviate from the mean more than $5\sigma$ before combining them. For each epoch in each aperture data set for $\tau$ Ceti, we average the calibration data points which have nearest epochs by weighting them according to their measurement errors. We further remove the outliers which deviate from the mean more than $3\sigma$. There are also epochs where no RVs of other stars are available, we assign the RVs measured at nearby epochs to them. We use these calibration data sets as proxies (like activity indices) to remove instrumental noise. For example, we can use a linear combination of the calibration data sets for C3AP2-1 and C3AP3-2 to model the instrumental noise in the C1AP1 data set. Hereafter, we use cC$n$AP$i$ (c$n$AP$i$) and cC$n$AP$i$-$j$ (c$n$AP$i$-$j$) to denote the calibration data sets for C$n$AP$i$ ($n$AP$i$) and C$n$AP$i$-$j$ ($n$AP$i$-$j$).

Through comparing various combinations of calibration data sets for all RV data sets, we find that the calibration data sets are appropriate proxies to reduce the instrumental noise in the data. Although the cC1AP1 data set is influenced by the instrument in the same way as the C1AP1 data set, the former is ``contaminated'' by Keplerian signals from other stellar systems though the contamination is reduced by removing outliers and averaging the data. On the contrary, the differential calibration data sets do not contain Keplerian signals, and thus are more appropriate for removing instrumental noise.
\begin{figure}\centering 
 \includegraphics[scale=0.65]{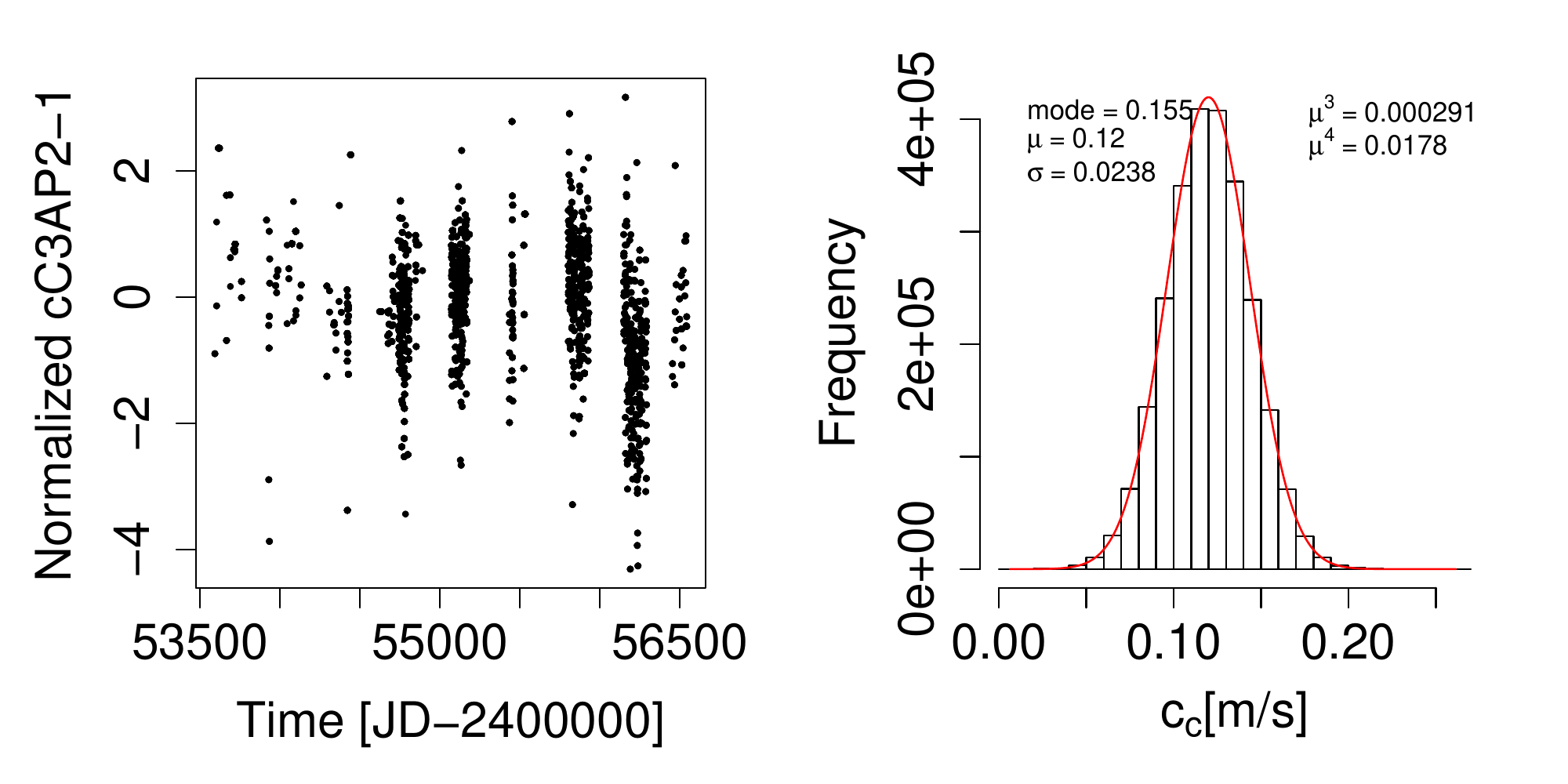}
 \caption{The normalized cC3AP2-1 calibration data set (left), and the posterior densities of the linear dependence of the C1AP1 data set on it (parameterized by $c_{c}$) for the 0P+MA(4)+8D model (right). The calibration data sets are normalized to zero mean and unit standard deviation before inclusion in the model. Thus $c_c$ is in m/s, characterizing the level of instrumental bias. The mode, mean, standard deviation and higher order moments of the posterior density are also shown. }
\label{fig:calibration}
\end{figure}
We find that the dependence of RV on calibration data sets is not consistent with zero. The HARPS measurements are biased at least by $0.244_{-0.035}^{+0.083}$\,m/s, determined by the square root of the sum of MAP estimations of linear coefficients for all cC9AP differential data sets. This bias is not reduced by including Keplerian components into the model, suggesting its instrumental origin. The real bias caused by the instrument could be higher because the differential calibration data sets only account for the wavelength-dependent instrumental noise. Thus the HARPS data would not be reliable for detecting signals below 0.2\,m/s if the instrumental noise is not properly removed. 

However, not all of the cC9AP differential data sets are useful, and the application of all of them only slightly improve the BF with respect to the cC3AP differential data sets. For the C3AP sets, only cC3AP2-1 is strongly correlated to the RV data, indicating higher instrumental noise in blue spectral orders than in red orders.
In Fig. \ref{fig:calibration}, we show the cC3AP2-1 data set for C1AP1, and the corresponding posterior distribution for the model. We see a correlation between cC3AP2-1 and C1AP1 up to 0.15\,m/s. Hence we will use the cC3AP2-1 data set to remove wavelength dependent instrumental noise in the HARPS data. Hereafter we include cC3AP2-1 together with the S-index, BIS and FWHM indices linearly in the model (see Eqs. \ref{eqn:full} and \ref{eqn:psi}) to analyze the C1AP1 data set. Similarly, we use c3AP2-1 to remove the instrumental noise in the 1AP1 and CCF data sets. We select the optimal number of MA components and differential RVs in the following subsection.

\subsection{Choosing the optimal noise model}\label{sec:opt_model}
Within the Bayesian framework, we compare noise models with various differential RVs and MA components to determine the optimal noise model for the C1AP1 data set. We report the Bayes factors with respect to the MA(0) model (without differential RVs or any Keplerian component) in Table \ref{tab:ap_comparison}.  The optimal number of differential RVs and MA components are selected based on the BIC-estimated Bayes factor threshold of 150 \citep{feng16}. In other words, the optimal model is the most complex model which gives a BF at least 150 times higher than all simpler models. This is equivalent to an increase of 5 of logarithmic BF in Table \ref{tab:ap_comparison}. 

\begin{table}
  \centering
\caption{BIC-estimated BFs of noise models (without Keplerian component) for the C1AP1 data set. The relative maximum logarithm likelihood for each model with respect to the white noise model (i.e. MA(0)) is shown. The BF of the Goldilocks noise model is shown in boldface.}
\label{tab:ap_comparison}
  \begin{tabular}{c|*{7}{c}}
  & MA(0) & MA(1) & MA(2) & MA(3) & MA(4) & MA(5) & MA(6) \\ 
  \hline
0D & 0.00 & 2234.55 & 2506.11 & 2606.66 & 2648.22 & 2649.77 & 2645.32 \\ 
  2D & 411.11 & 2576.22 & 2840.77 & 2961.32 & 3013.88 & 3016.43 & 3011.98 \\ 
  5D & 504.77 & 2596.88 & 2868.43 & 2989.98 & 3041.54 & 3041.09 & 3039.65 \\ 
  8D & 537.43 & 2609.54 & 2883.09 & 3005.65 & {\bf 3053.20} & 3056.75 & 3053.31 \\ 
  17D & 548.41 & 2596.52 & 2867.08 & 2994.63 & 3043.18 & 3044.74 & 3041.29 \\ 
 \end{tabular}
\end{table}

\begin{figure}\centering 
  \includegraphics[scale=0.8]{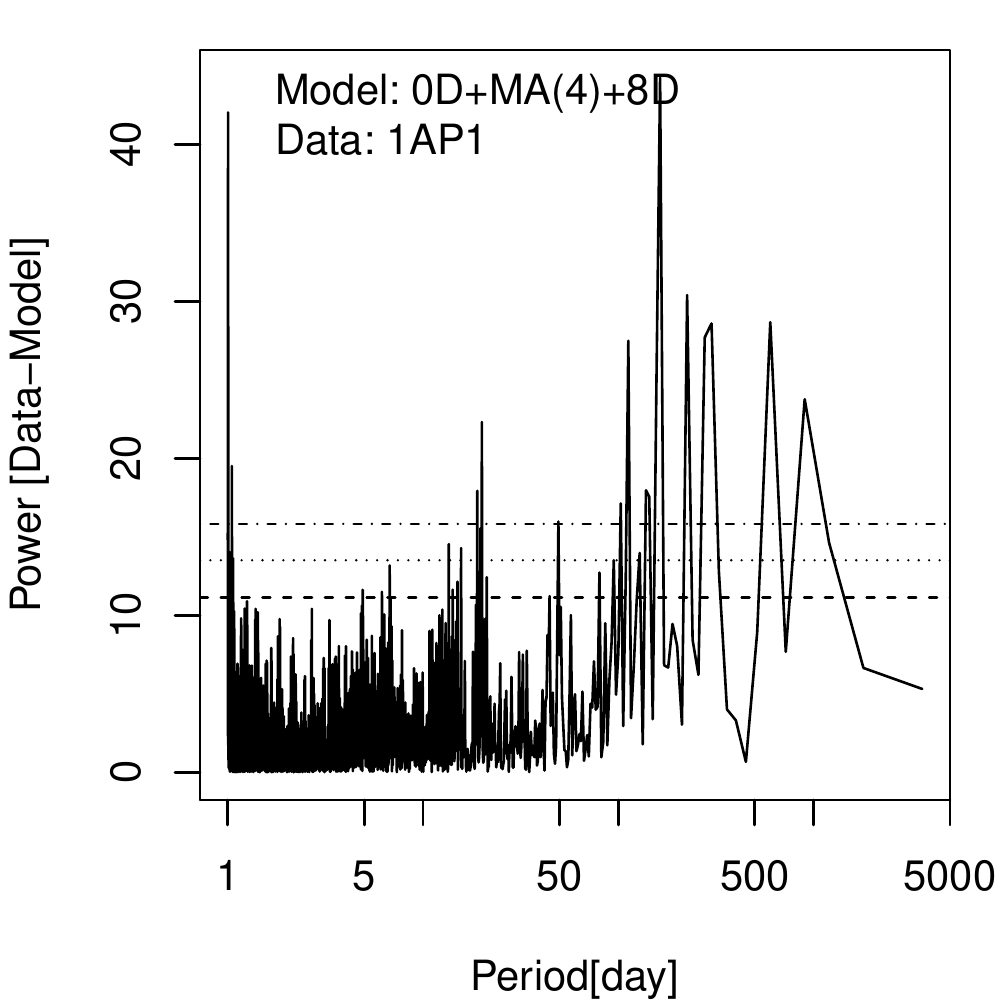}
  \includegraphics[scale=0.8]{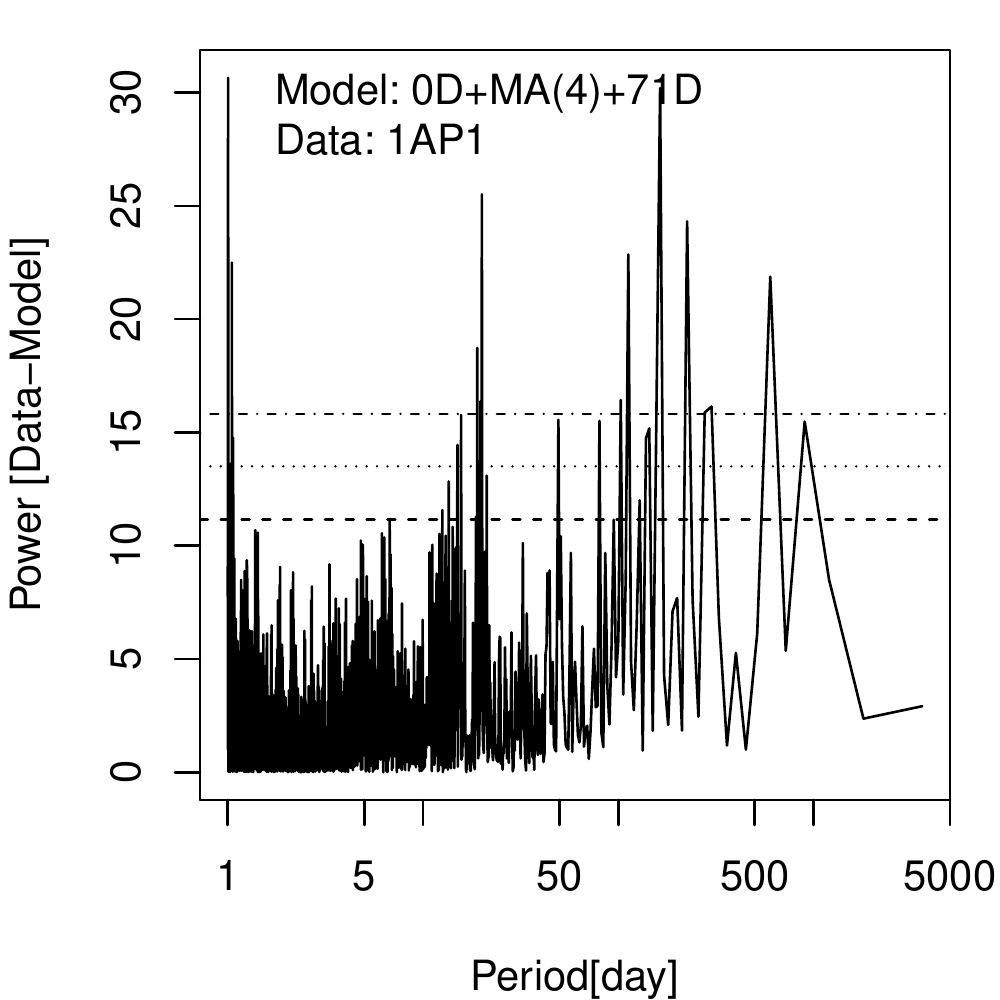}
  \caption{The comparison between the periodograms for the residuals of the 1AP1 data set after subtracting the 0D+MA(4)+8D (left) and 0D+MA(4)+71D (right) prediction. The horizontal lines denote the FAPs of 0.1, 0.01 and 0.001.} 
\label{fig:goldilocks_ap}
\end{figure}

According to this criterion, the 0P+MA(4)+8D model is the noise model favored by the 1AP1 and C1AP1 data sets. Including more differential RVs may not reduce the noise, rather, the noise in the differential RVs may reduce the significance of true signals. To explain this we compare the periodograms for the residuals of the 1AP1 data set after subtracting the 0P+MA(4)+8D and 0P+MA(4)+71D models in Fig. \ref{fig:goldilocks_ap}. Guided by the power difference between FAP thresholds in the periodograms, we see that the noise in the residuals increases if including more differential RVs. Moreover, complex noise models may interpret signals as noise due to their flexibility (\citealt{feng16} and MT13). Based on the above considerations, we adopt the 0P+MA(4)+8D model to remove RV noise. We also find at most weak correlation between noise proxies (see section \ref{sec:correlation}), supporting the necessity of using all of them in noise modeling. This Goldilocks model is different from the one devised by \cite{feng16} for M dwarfs because the target and data sets in this work are different. But results both in this work and in \cite{feng16} suggest the importance of finding an appropriate noise model for each specific RV data set. 

Following this approach, we further find the Goldilocks noise models for the CCF data set to be 0P+MA(5)+8D. The optimal noise model for the KECK data set is 0P+MA(1)+0D. In the 0P+MA(4)+8D model, there are 2 free parameters for trend, 1 parameter for jitter, 4 parameters for activity indices and c3AP2-1, 5 parameters for the MA model and 8 parameters for differential RVs. Thus there are 20 and 21 free parameters in the 0P+MA(4)+8D and 0P+MA(5)+8D models, respectively. Since the differential RVs can weight aperture data sets {\it a posteriori} and thus remove wavelength-dependent noise (see section \ref{sec:weight}), it is unnecessary to analyze aperture data sets separately. Hence we only analyze the averaged data sets,  C1AP1, 1AP1 and CCF, to identify signals. 

\subsection{Performance of the Goldilocks noise model}
\begin{figure}
  \centering 
\includegraphics[scale=0.5]{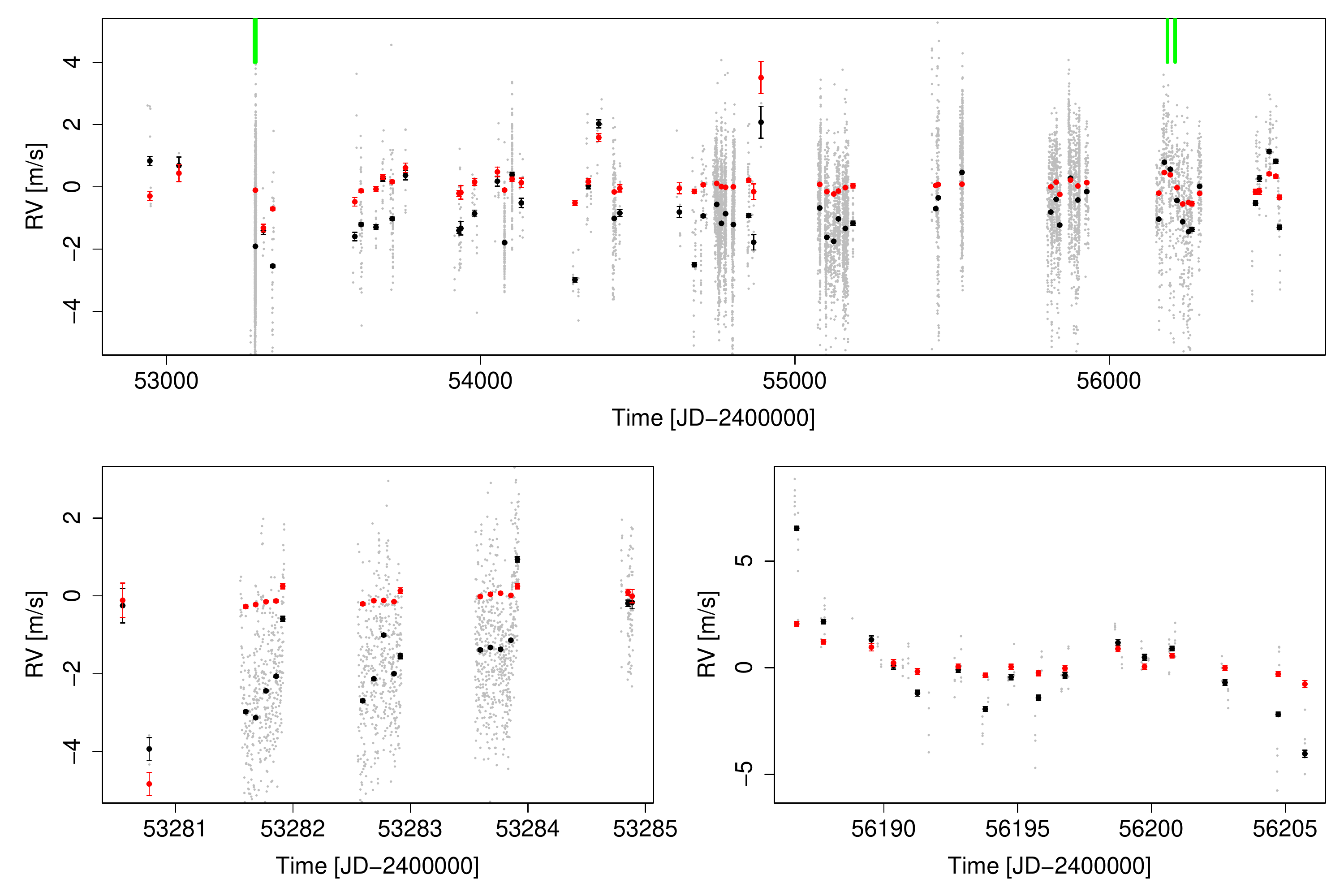}
\vspace*{0.1in}
  \caption{The RVs of all observation epochs (top) and for epochs around JD2453283 (bottom left) and JD2456195 (bottom right) for the 1AP1 data set. The bottom two panels show the RVs measured in the time spans denoted by the green lines in the top panel. The black and red dots with error bars represent binned RVs and binned residuals after subtracting the best-fitting noise model, respectively. }
  \label{fig:2epochs}
\end{figure}

Based on visual investigation of the 1AP1 data set, we see a rapid increase in RV around epoch JD2453282 and decrease around epoch JD2456190, as shown by black points in Fig. \ref{fig:2epochs}. We see a steady increase of RV by 5\,m/s and a rapid decrease by 10\,m/s around the above epochs. These variations are huge compared with the sub-meter semi-amplitudes of the signals which we will report in the following section. Any model that does not fit these two features would have low likelihood, and thus not be favored by the data. 

We use the Goldilocks noise model, 0P+MA(4)+8D, to model the noise in the 1AP1 data set and show the binned residual after subtracting the best-fitting model as red points in Fig. \ref{fig:2epochs}. We see that the noise model significantly reduce the noise-induced RV variation. However, we still see weak intra-night noise in the epochs around JD2453283, probably caused by the guiding error, as mentioned in section \ref{sec:data}. Hence the C1AP1 data set is probably a more conservative choice for signal detection. According to our analysis, the KECK data does not help to constrain the signals detected in HARPS data sets. To be conservative, we will analyze C1AP1 and CCF to identify signals, and use KECK and 1AP1 for sensitivity and consistency test in the following sections. 

\section{Keplerian signals}\label{sec:signals}
In the above section, we obtain the optimal noise model for the TERRA-reduced HARPS data set. Since the spectral orders of the CCF data is not available, we apply the TERRA differential RVs to model the RV noise in the CCF data set \footnote{Although these differential RVs are produced by TERRA, they should be able to remove wavelength-dependent noise in the CCF data since the two data sets are rather similar, leading to at most a second order difference in the differential RVs produced by TERRA and CCF. Even though they produce rather different differential RVs, we can still regard TERRA differential RVs as a type of activity indices and consider their linear correlation with the CCF RVs. }. From now on, we only use the C1AP1 data set to identify signals and use 1AP1 and CCF to test the consistency of signals. 

\subsection{Primary signals}\label{sec:primary}
\begin{figure}[h!]
  \centering 
      \includegraphics[scale=0.55]{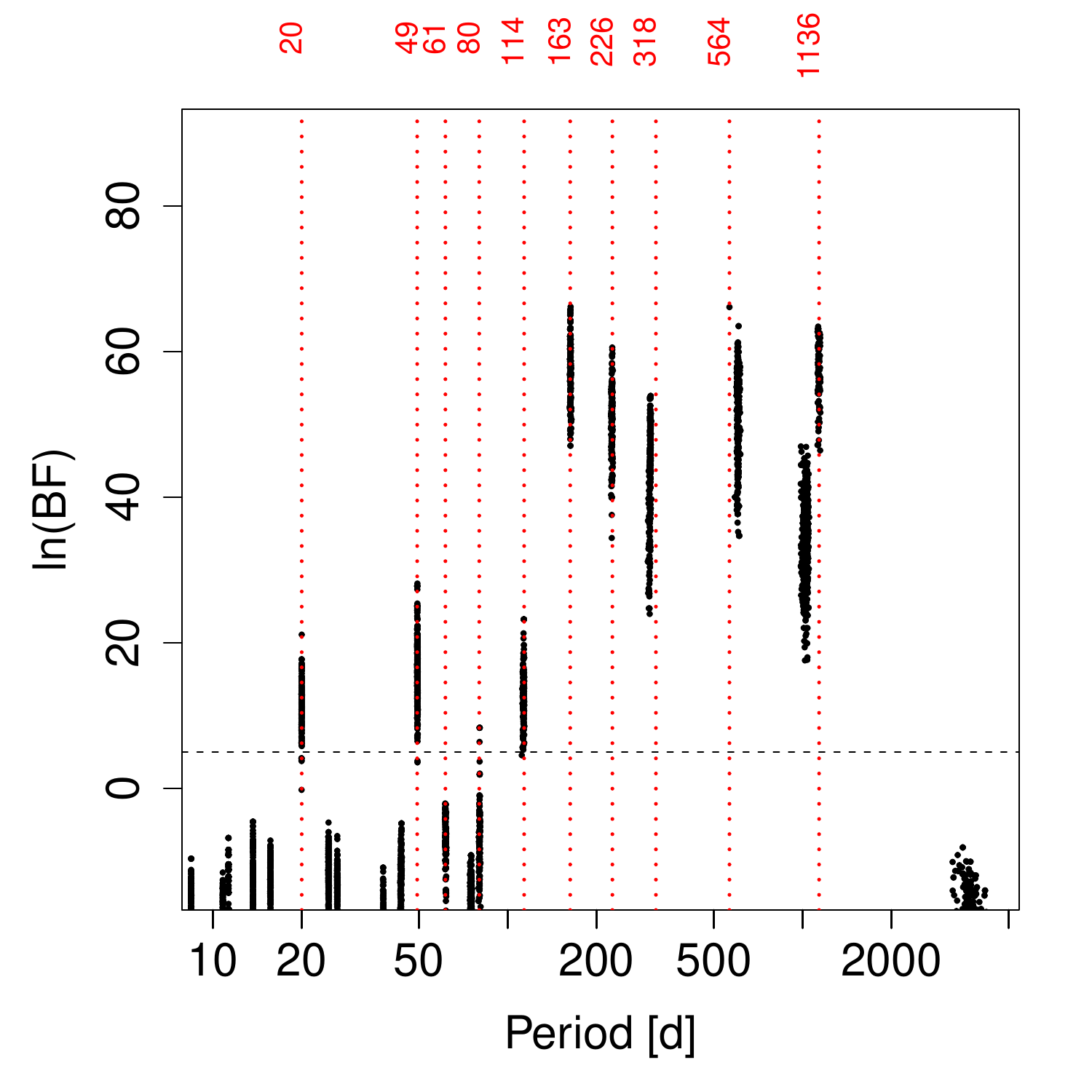}
    \includegraphics[scale=0.55]{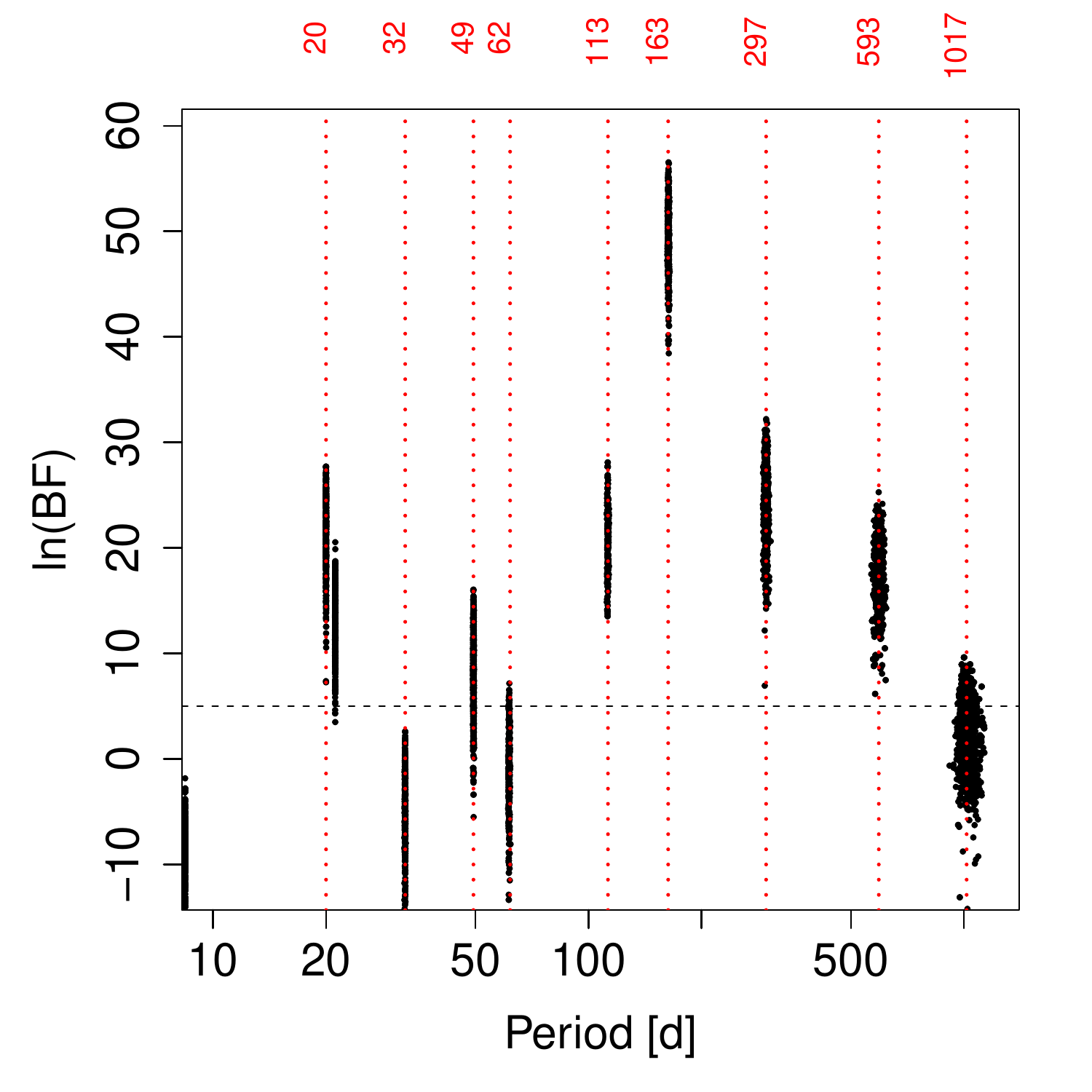}
  \caption{The logarithmic BF distributions of the samples drawn by the cold chains for the Keplerian (left) and circular (right) solutions of the 1P+MA(4)+8D model for the C1AP1 data set. There are 10,000 samples randomly drawn from the each cold chain. The periods of some local maxima are denoted by red lines and numbers. If two maxima are close to each other, we only show the one with higher posterior. The BIC-estimated logarithmic BF threshold of 5 is denoted by the dashed line in each panel. }
\label{fig:primary}
\end{figure}

As is mentioned in section \ref{sec:method}, we run hot chains to find primary signals for later investigations. We divide the period range into 20 chunks and run a hot chain to find the local posterior maxima for each. We convert the tempered posterior to the posterior for each chain, and combine all posteriors to approximate the posterior distribution for the 1P+MA(4)+8D model. We also start cold chains from the local posterior maxima of hot chains, and combine the posteriors drawn by cold chains. To compare with signals detected in the Keplerian solution, we also fit sinusoidal functions to the data to obtain circular solutions. In Fig. \ref{fig:primary}, we show the distributions of logarithmic BF (estimated by BIC at a given period) over period for the Keplerian and circular solutions for the 1AP1 data set. We see that the strongest signals are around 160\,d, 600\,d and 1000\,d. The 600 and 1000\,d signals are probably annual aliases of each other. The 114\,d and 318\,d signals are annual aliases of 160\,d while 226\,d is an annual alias of 600\,d. The signals at periods of about 20, 49, 160, 600 and 1000\,d and some of their annual aliases are also significant in the Keplerian solution. 

\begin{figure}[h!]
  \centering 
  \includegraphics[scale=0.80]{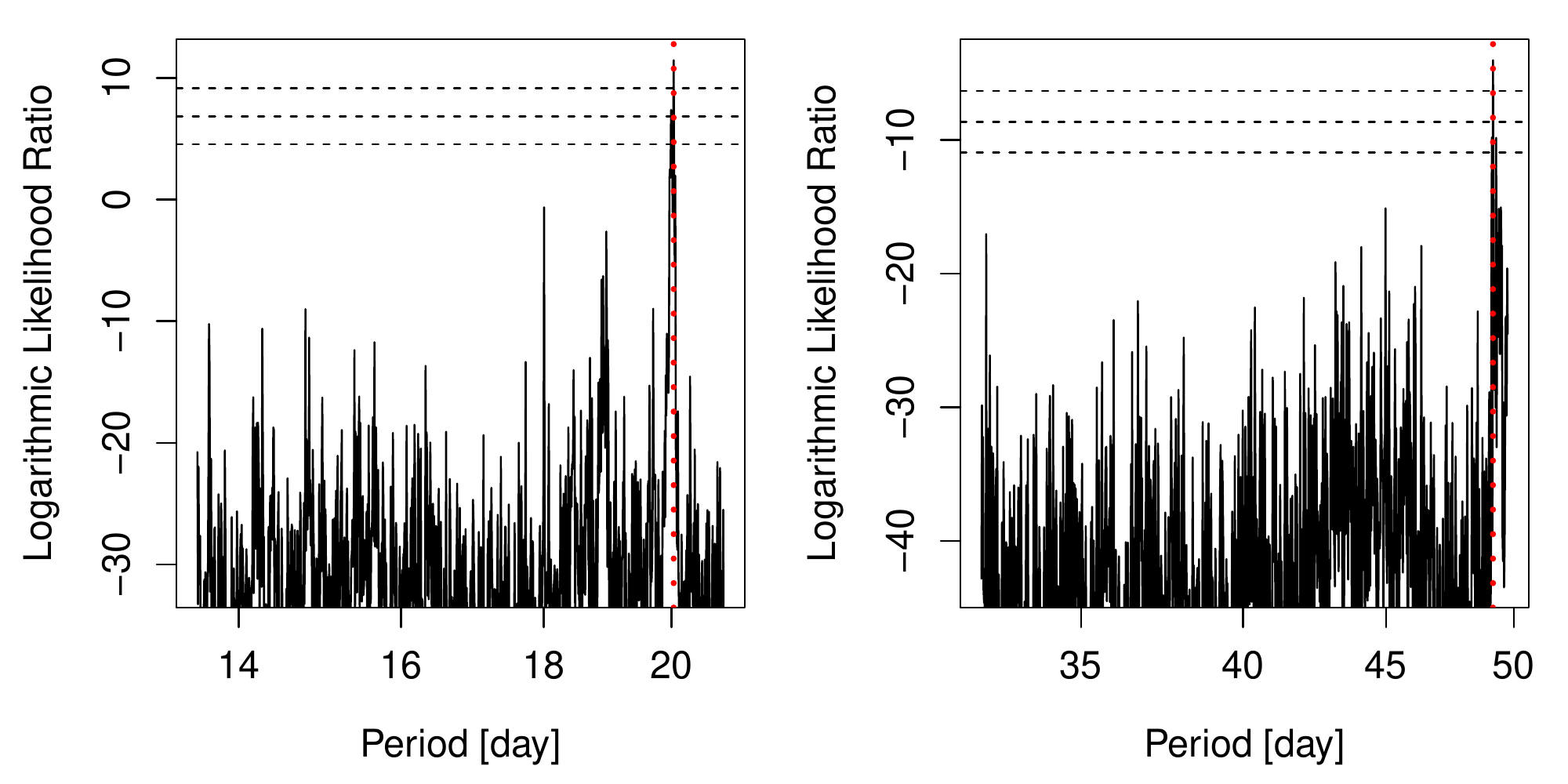}
  \caption{The logarithmic likelihood ratio of the 1P+MA(4)+8D model and the 0P+MA(4)+8D model based on the 3.8 million posterior samples drawn by hot chains for the period intervals of $[13.5,20.9]$\,d (left) and $[32.24,49.76]$\,d (right) for the C1AP1 data set. Each period range is divided into 1000 bins, and the maximum posterior is determined for each bin. The posterior distributions shown here are linear interpolations of these posteriors. The 10\%, 1\% and 0.1\% quantiles of the maximum logarithmic likelihood ratio are denoted by dashed lines. Note that the likelihood ratio determined by hot chains does not represent the real one.}
\label{fig:2post}
\end{figure}

To demonstrate the uniqueness of the primary signals, we show the posterior distribution for two period ranges where the signals around 20\,d  and 49\,d are identified in Fig. \ref{fig:2post}. We see that the two signals are unique and well identified by the hot chains in the corresponding period intervals. Therefore it is reliable to use the parameters of these signals as initial conditions for cold chains to constrain signals simultaneously. 

\begin{figure}
  \centering 
    \includegraphics[scale=0.50]{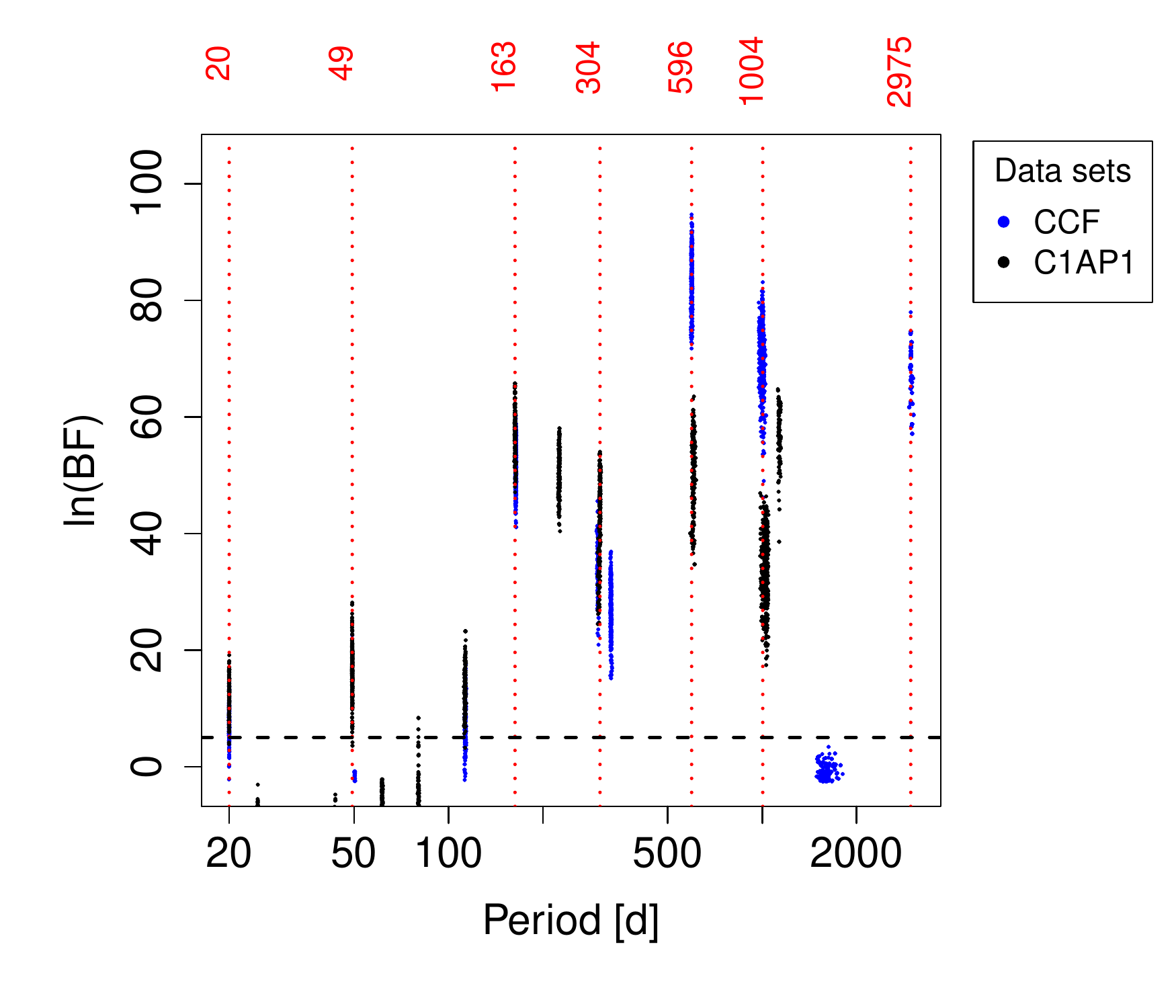}
  \includegraphics[scale=0.50]{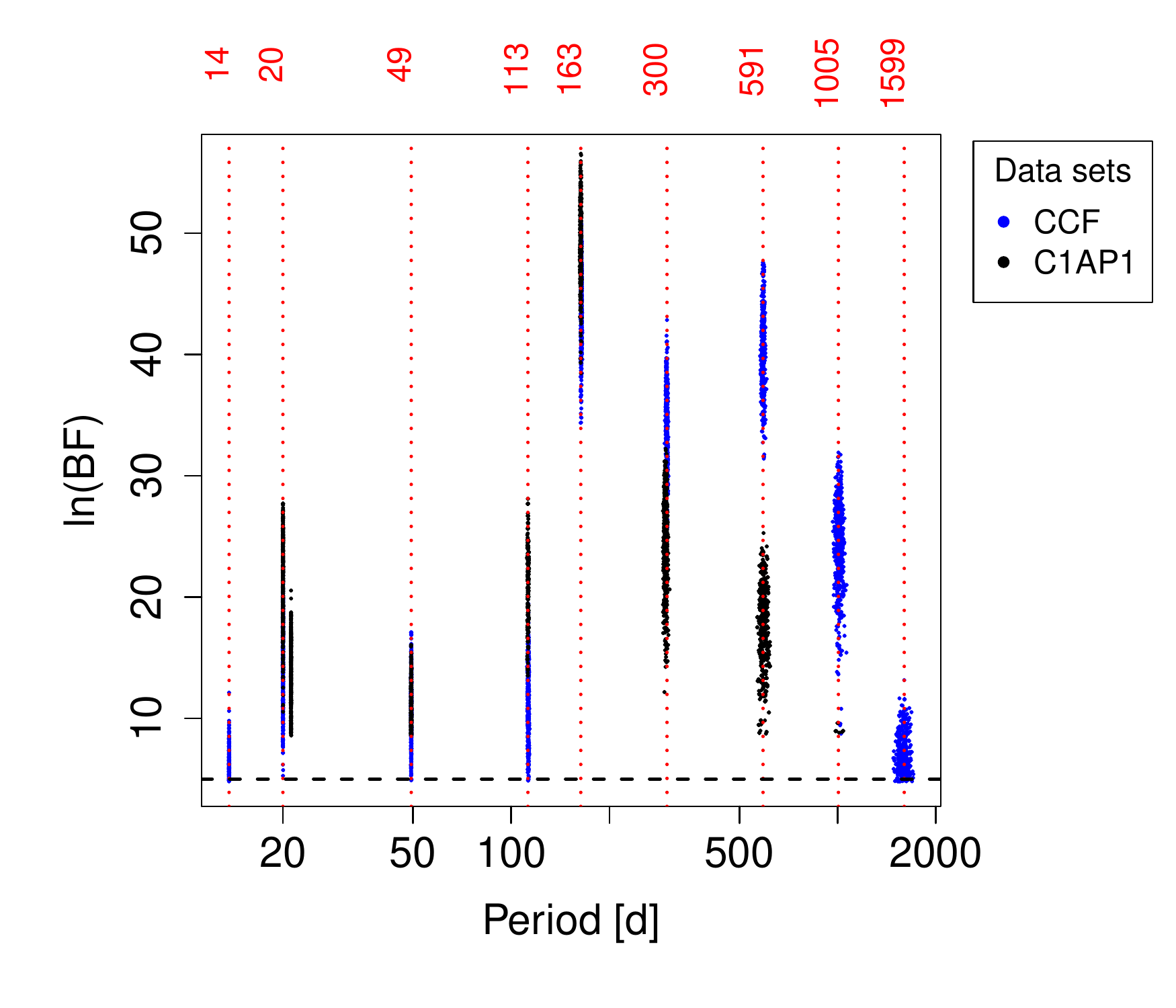}
  \caption{Similar to Fig. \ref{fig:primary}, but for cold chains for the C1AP1 and CCF data sets.}
\label{fig:all_peaks}
\end{figure}

In Fig. \ref{fig:all_peaks}, we show the logarithmic BF distributions of the samples drawn by cold chains for the C1AP1 and CCF data sets. For both solutions and data sets, the signals around 160\,d, 600\,d and 1000\,d are most significant. The 600\,d signal is more significant than the 1000\,d signal for both Keplerian and circular solutions for both data sets. According to our analysis, the signals at periods of 20, 49, 160, 600 and 1000\,d can also be found in the 1AP1 data set and combinations of HARPS and KECK data sets. However, the comparison of the significance of primary signals is biased because the one-planet model may interprets the variations caused by multiple signals as one Keplerian signal. Thus we will constrain these primary signals simultaneously, and compare the results for various data sets in the following section. 

\subsection{Comparing signals detected in different data sets}\label{sec:comparing}
Using the numerical method described in section \ref{sec:method}, we obtain posterior samples from the MCMC sampling for RV models with various numbers of Keplerian components. We select signals according to the signal detection criteria within the Bayesian framework. The parameters of signals identified in CCF and C1AP1 are shown in Table \ref{tab:signals}, and the BFs for the Keplerian and circular solutions for the C1AP1 set are shown in Table \ref{tab:BFs}. The signals detected in the KECK data set are not shown because only a signal at 20\,d is identified. The strongest signals in Keplerian solutions for both data sets have orbital periods around 1000\,d and 600\,d. Since they are annual aliases, we report the results for both in the table. We find that the eccentricities of these two signals are high for CCF but are below 0.2 for C1AP1. This indicates that the high eccentricity is probably caused by the noise in the observation epochs before JD2453500 (see Fig. \ref{fig:data}) which are included in CCF but not in C1AP1.

\begin{table}
  \centering
  \caption{The MAP estimation of the parameters of signals for the CCF and C1AP1 data sets and for both Keplerian (left columns) and circular solutions (right columns). The parameters shown are the period ($P$ in days), semi-amplitude ($K$ in m/s) and eccentricity ($e$) of the signals. For circular solutions, only semi-amplitude and period are shown. For each signal, the parameters are estimated in combination with the $\sim$1000\,d and with the $\sim$600\,d signals are shown in upper and lower entry respectively for each parameter and data set. The solutions are obtained based on at least 4 million MCMC samples.}
  \label{tab:signals}
\begin{tabular}  {c*{2}{c}|*{2}{c}}
  \hline
&  \multicolumn{2}{c|}{Keplerian}&  \multicolumn{2}{c}{Circular}\\
&CCF &  C1AP1& CCF&C1AP1\\\hline
  \multirow{2}{*}{$P_1$}&1007.66&995.78&1015.25&942.68\\
&597.02 &636.13&598.55&619.39\\\hline
\multirow{2}{*}{$K_1$}&0.70&0.39&0.30&0.26\\
                &0.68&0.35&0.43&0.36\\\hline
  \multirow{2}{*}{$e_1$}&0.46&0.30&--&--\\
                & 0.58&0.16&--&--\\\hline
\multirow{2}{*}{$P_2$}&164.38&161.40&164.28&162.99\\
               &164.22&162.87&164.72& 163.42\\\hline
\multirow{2}{*}{$K_2$}&0.53&4.65&0.44&0.49\\
&0.48&0.55&0.47&0.51\\\hline
\multirow{2}{*}{$e_2$}&0.26&0.96&--&--\\
&0.18&0.18 &--&--\\\hline
\multirow{2}{*}{$P_3$}&20.03&20.03&20.03&20.01\\
                & 20.04&20.00&20.03&20.03\\\hline
\multirow{2}{*}{$K_3$}&0.47&0.42&0.41&0.46\\
                &0.49 &0.49&0.39&0.43\\\hline
\multirow{2}{*}{$e_3$}&0.18&0.11&--&--\\
                & 0.25&0.06&--&--\\\hline
\multirow{2}{*}{$P_4$}&49.47&49.29&49.36&49.37\\
                & 49.38&49.41&49.48&49.50\\\hline 
\multirow{2}{*}{$K_4$}&0.45&0.48&0.38&0.32\\
                & 0.42&0.39&0.42&0.42\\\hline 
\multirow{2}{*}{$e_4$}&0.28&0.29&--&--\\
                & 0.22&0.23&--&--\\\hline 
\multirow{2}{*}{$P_5$}&91.59&102.43&92.29&91.97\\
&91.97&91.34&91.65&91.72\\\hline
\multirow{2}{*}{$K_5$}&0.52&0.46&0.27&0.28\\
&0.32&0.45&0.32&0.30\\\hline 
\multirow{2}{*}{$e_5$}&0.53&0.15&--&--\\
&0.27&0.40&--&--\\\hline 
\multirow{2}{*}{$P_6$}&--&--&6.63&--\\
&--&--&6.63&31.68\\\hline 
\multirow{2}{*}{$K_6$}&--&--&0.28&--\\
&--&--&0.25&0.30\\\hline 
\multirow{2}{*}{$e_6$}&--&--&--&--\\
&--&--&--&--\\\hline
  \multirow{2}{*}{$P_7$}&--&--&14.21&--\\
&--&--&14.22&--\\\hline 
\multirow{2}{*}{$K_7$}&--&--&0.27&--\\
&--&--&0.19&--\\\hline 
\multirow{2}{*}{$e_7$}&--&--&--&--\\
&--&--&--&--\\\hline 
\end{tabular}
\end{table}

\begin{figure}\centering 
\includegraphics[scale=0.4]{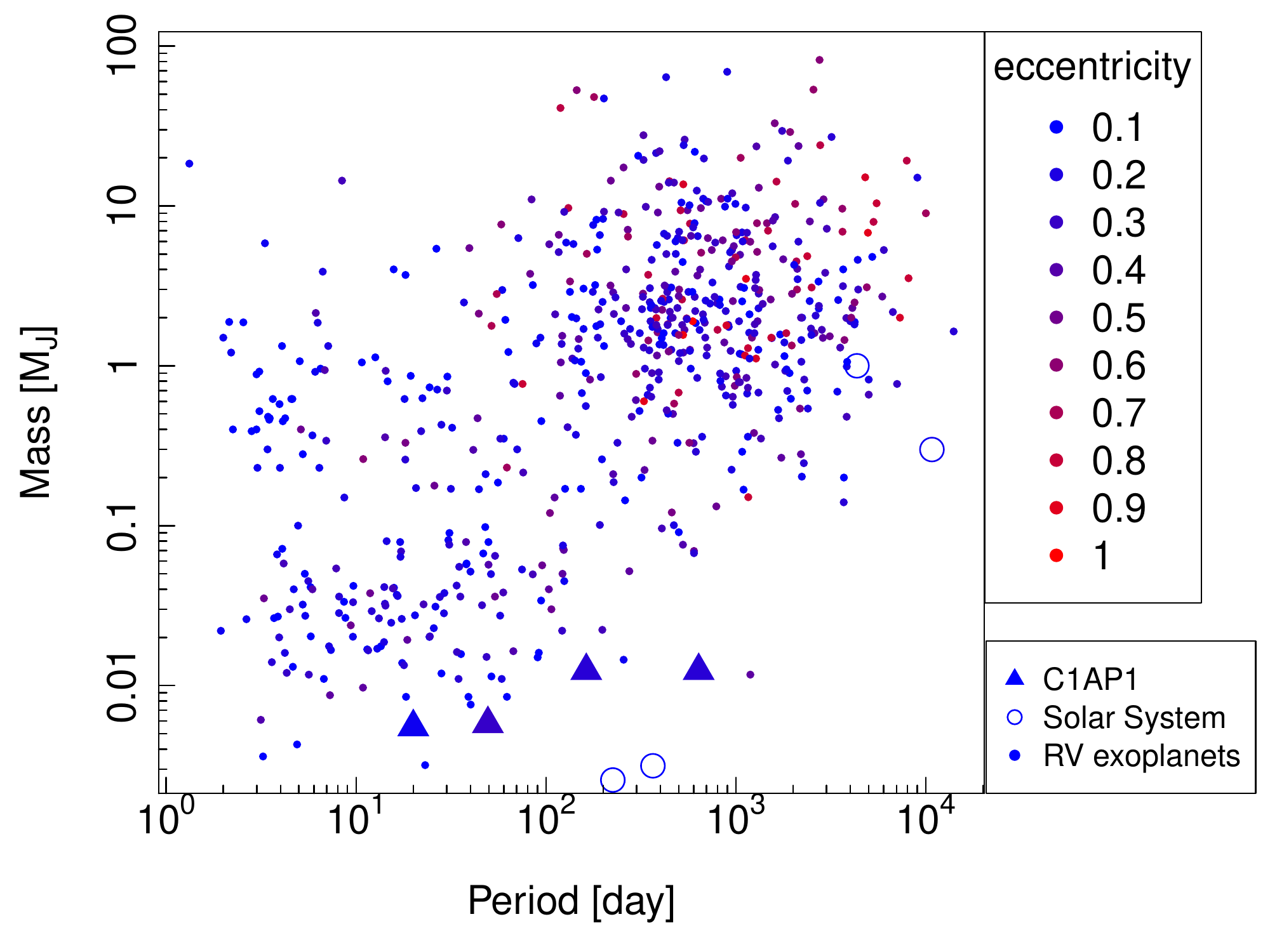}
\caption{Distribution of orbital periods and masses of the candidate planets detected in the C1AP1 data set (triangles) and all exoplanets detected by the radial velocity technique (small dots). To compare the $\tau$ Ceti candidate planets with Solar System planets, the parameters of the Venus, Earth, Jupiter and Saturn are shown in open circles from left to right. The eccentricity is encoded in the color of the points. The data are collected from the Extrasolar Planets Encyclopaedia (http://exoplanet.eu).}
\label{fig:RV_planets}
\end{figure}

\begin{table}
\caption{The logarithm BFs estimated by the BIC for models with various numbers of planets with respect to the noise model. The signals are identified in the Keplerian and circular solutions for the C1AP1 data set. The period of the long period signal is shown in brackets. Note that the logarithm BF threshold is 5. }
\label{tab:BFs}
\centering
\begin{tabular} {l*{6}{c}}
\hline  
 Number of planets&1&2&3&4&5&6\\\hline
Keplerian solution ($\sim$1000\,d)&46.6 &89.7& 191& 219&--&--\\ 
Keplerian solution ($\sim$600\,d)&63.7 &85.3 &126 &152& --&--\\
Circular solution ($\sim$1000\,d)&11.6 &59.1 &95.3& 203 &112&-- \\ 
Circular solution ($\sim$600\,d)&26.8 &72.6 &104 &118 &125 &134 \\
\hline
\end{tabular}
\end{table}
Highly eccentric and low mass exoplanets are very rare \citep{tuomi13}, which is evident from the distribution over the mass and eccentricity of all exoplanets detected through the RV technique in Fig. \ref{fig:RV_planets}. Although there are many biases in a plot like this and particularly at low masses and high eccentricities, we do see a lack of planets with low mass, long period and high eccentricity like the 1000\,d signal we have identified in this work. Thus the high eccentricity solutions reported in Table \ref{tab:signals} are probably not caused completely by planets orbiting the star but are superpositions of planets and activity. The activity and signals are not completely disentangled probably because of incomplete noise modeling or inefficiency of the MCMC sampling, which makes the chain unable to jump out of local maxima to find extra low eccentric signals. We investigate the former by adding more MA components to the noise model, but fail to reduce the eccentricity. Then we investigate the latter by finding signals with zero eccentricity, which makes the MCMC chains achieve convergence more efficiently due to reduction in dimensionality. Based on the Bayesian model comparison, the circular solutions identify more signals than the Keplerian solutions because the sinusoidal function has less parameters than the Keplerian function and thus is less penalized by the BIC. We find consistent circular solutions of five signals for all data sets. The 92 and 102\,d signals are probably an alias pair since the subtraction of one from the data would weaken the other. The 92\,d signal may be genuine because it is identified in circular solutions as well as the Keplerian solution for CCF. Notably the posterior of the 92\,d signal is rather low in the Keplerian and circular solutions because it does not pass the BF threshold as a primary signal, as shown in Fig. \ref{fig:all_peaks}. 

For both of the circular and Keplerian solutions, we find that the 600\,d signal is favored over the 1000\,d signal for all data sets. This also leads to a slightly higher amplitude of the 600\,d signal with respect to the 1000\,d signal (see Table \ref{tab:signals}). Moreover, the 160\,d signal is very eccentric ($e=0.96$) in the 1000\,d solution for C1AP1. Thus the long period signal probably has a period around 600\,d, if it is caused by a planet. This signal could be confirmed by further observations of the inner edge of the debris disk of $\tau$ Ceti by the Atacama Large Millimeter/sub-millimeter Array (ALMA) \citep{macgregor16}. If the inner edge is found to be beyond 1.5\,AU, this planet must exist to clear the inner disk. As for the other signals, we regard the 20\,d signal as a genuine planetary candidate since it is also identified independently in the KECK data set. The signal around 160\,d is Keplerian because it is the most significant signal in the circular solutions and is consistently identified in all HARPS data sets. The signal at a period of 49\,d is Keplerian because it is identified in all data sets. There could be a signal at a period of about 92\,d or 102\,d. But the confirmation of this requires further observations and analysis. These results are also valid for the 1AP1 data set and combinations of HARPS and KECK data sets according to our analysis. 

\subsection{Curse of eccentricity}\label{sec:curse}
Although the eccentricity of the 600\,d signal is low for the C1AP1 data set, the other signals have eccentricities which are not consistent with zero. We also add extra Keplerian signals to the Keplerian solutions for the C1AP1 and CCF data sets, but the high eccentricity of signals persists. To decide whether the signals are time dependent, we evenly divide the time span of the 1AP1 data set into three chunks, and obtain Keplerian solutions for each. We find that the jitter level for the first chunk is about 0.3\,m/s higher than the other two chunks. The jitter may significantly dilute the signal in the first chunk because they have comparable semi-amplitude which is around 1\,m/s. The 1000\,d signal is identified in the second and third chunks, although it is rather eccentric in the third chunk. The 600\,d signal is identified in the third chunk. The 49 and 160\,d signals are identified in the third chunk. The 20\,d signal is found in all chunks while the 14\,d signal is only significant in the first chunk. This is probably the reason why MT13 has identified the 14\,d signal rather than the 20\,d one based on analyses of early data. This apparent inconsistency of signals is unsurprising because they are signals with semi-amplitudes of at most 0.5\,m/s and require large samples to confirm. Although chosen based on analyses of the whole HARPS data, the 1P+MA(4)+8D model is applied here to different chunks, which probably lead to false negatives.

 We investigate this by dividing the 1AP1 data into 3 chunks, applying the 0P+MA(2)+2D noise model to each, and constraining signals using the whole data set. In the periodogram for the residual after subtracting the noise model prediction, we are able to identify all signals despite the noise parameters varying with chunks. This suggests that the signals we have identified are probably not caused by activity cycles. 

\subsection{Instrumental bias of HARPS}\label{sec:instrument}
The above investigations of the cause of high eccentricity are concerned with the removal of activity-induced noise from the data. In this section, we will study the excess instrumental noise caused by the instability of HARPS in the sub-m/s regime. We model this noise using the moments of the HARPS line profiles, which probably reflect the flux loss caused by instrumental effects such as guiding errors \citep{berdinas16}. Following \cite{berdinas16}, the moments are calculated in the following steps:
\begin{itemize}
\item The fluxes of all spectra are scaled such that the relative flux in each echelle order is the same.
\item The spectrum recorded in all echelle orders is deconvolved to obtain a mean line profile $F_i(t)$ in absorption. After subtracting the residual continuum $w(t)$, the line profile is inverted and normalized to be a flux distribution represented by $N$ flux values $f_i(t)$ and RVs $v_i(t)$. 
\item The moments are calculated according to the following equations,
\end{itemize}
\begin{equation}
  M_n(t)=\dfrac{\sum_{i=1}^{N}f_i(t)v_i(t)^n}{\sum_{i=1}^{N}f_i(t)}~, \\
\label{eqn:moments}
\end{equation}
where
\begin{equation}
f_i(t)=w(t)-F_i(t)~,
\end{equation}
where $n$ is a natural number. The zero moment is simply the sum of flux values, $M_0=\sum_{i=1}^{N}f_i(t)$. 
The central moments are defined by 
\begin{equation}
  M_{nc}(t)=\dfrac{\sum_{i=1}^{N}f_i(t)[v_i(t)-M_1(t)]^n}{M_0(t)}~. \\
\label{eqn:central_moment}
\end{equation}
To minimize excess instrumental noise, we use the central moments together with $M_0$ because the non-central moments contain a certain amount of information from Keplerian signals. Although the central moments attempt to minimize the mean velocity contribution by subtracting $M_1$, but $M_1$ is not perfect at removing Keplerian velocities. So we only use central moments to investigate the instrumental noise rather than to quantify signals before any robust tests on the connection between moments and RVs being done. 

\begin{figure}\centering 
\includegraphics[scale=0.6]{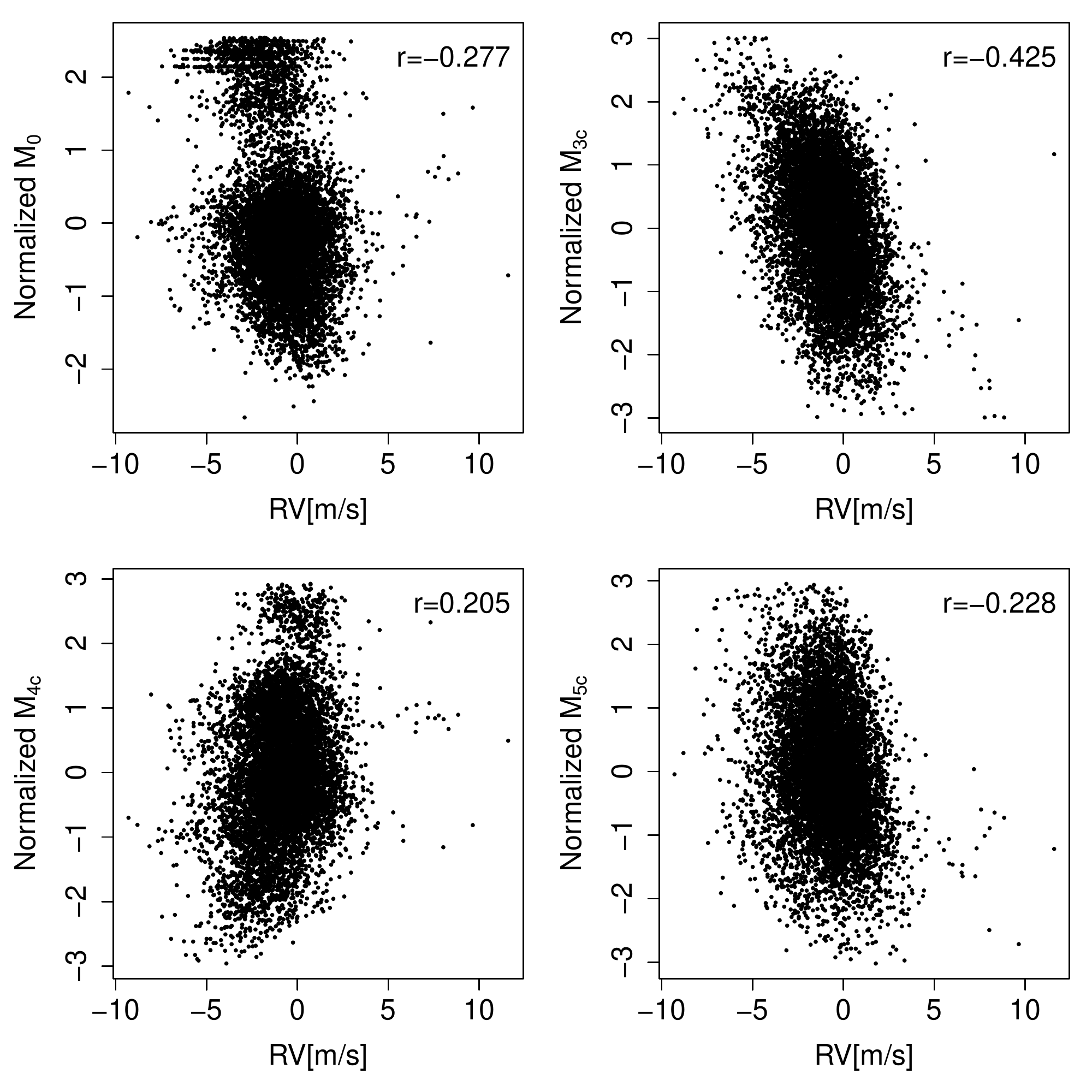}
  \caption{The correlation between RVs and $M_0$, $M_{3c}$, $M_{4c}$and $M_{5c}$ for the C1AP1 data set. The Pearson correlation coefficient is shown in the top right corner. }
\label{fig:moments}
\end{figure}

We linearly combine $M_0$, $M_{3c}$, $M_{4c}$ and $M_{5c}$\footnote{We don't use $M_{2c}$ because it is equivalent to FWHM.}, activity indicators, calibration data sets and differential RVs in the model, and find strong correlations between RVs and these moments. For example, we use this new noise model to constrain the 160\,d signal which is identified in the C1AP1 data set. We find that the linear dependence of RVs on the normalized $M_0$, $M_{3c}$, $M_{4c}$ and $M_{5c}$ are around $-0.17\pm0.02$\,m/s, $-0.62\pm0.02$\,m/s, $0.16\pm 0.02$\,m/s and $-0.28\pm 0.02$\,m/s, respectively\footnote{These values are the mean and standard deviation of the posterior samples drawn by cold chains, and thus are different from the Pearson coefficients shown in Fig. \ref{fig:moments}.}. In particular, these moments increase the logarithm BF by more than 1000 for all data sets. We show this strong correlation between RVs and moments for the C1AP1 data set in Fig. \ref{fig:moments}. We observe strong anti-correlations between RVs and odd order moments which have terms proportional to $-M_1$. This indicates that the subtraction of $M_1$ from velocities (see Eqn. \ref{eqn:central_moment}) may not completely remove the Keplerian components, leading to a considerable correlation between RVs and central moments. We apply the four moments to all data sets by including them into the model linearly. We find that the strong signals shown in Table \ref{tab:signals} are still significant with this noise model. 
 
\begin{figure}
  \centering 
  \includegraphics[scale=0.7]{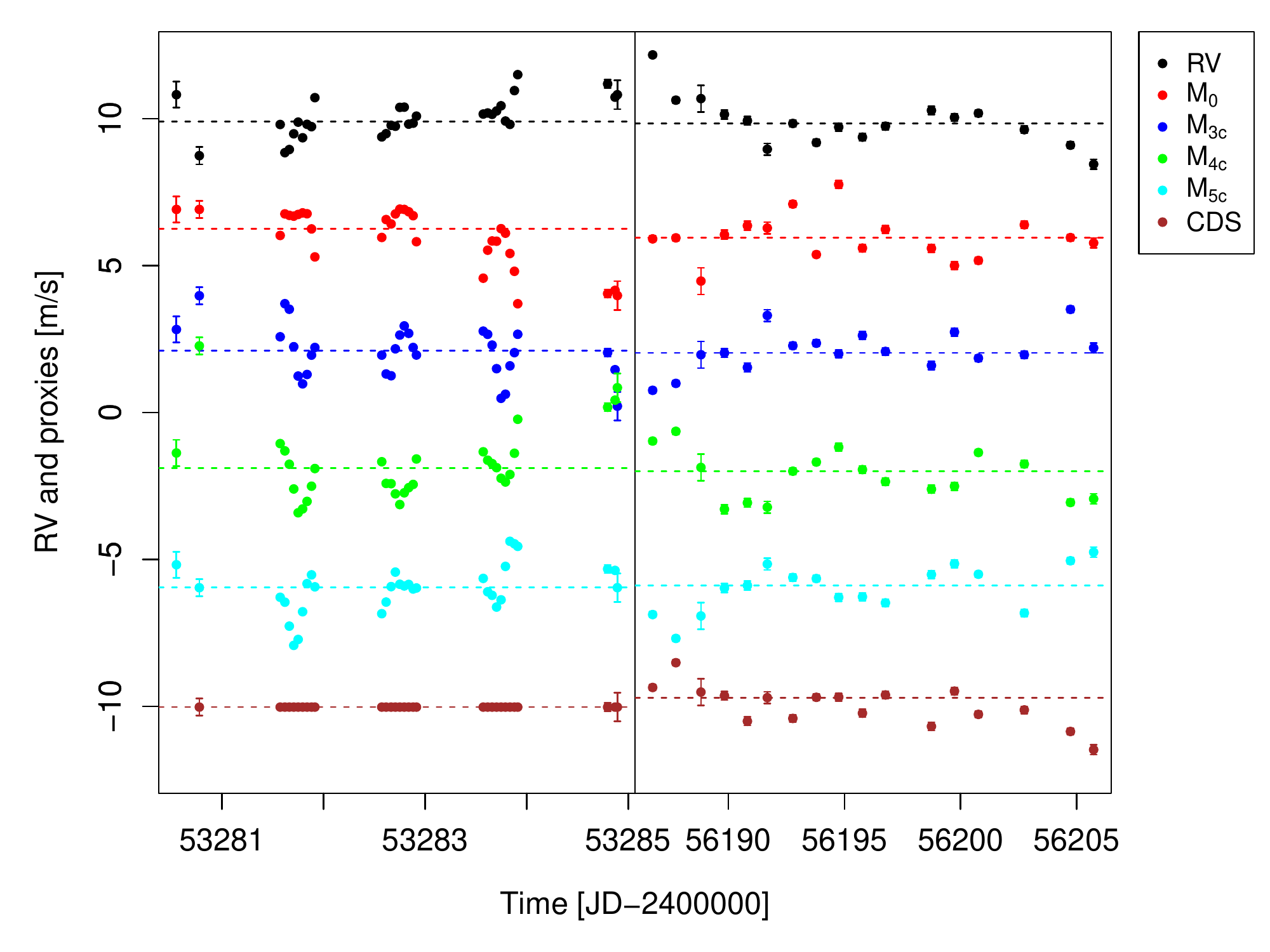}
  \caption{The normalized binned RVs, moments and c3AP2-1 data set (CDS) around epochs of JD2453283 and JD2456195 for the 1AP1 data. The binning windows for the left and right panels are 0.05\,d and 0.5\,d, respectively. The c3AP3-2 data set around JD2453283 are set to the value of nearby epochs since no calibration stars are available  around this epoch. The mean of each time series is shown by horizontal dashed line  while all time series are shifted up/down for visualization.}
\label{fig:moments_2epochs}
\end{figure}

To explore the correlation between the moments and calibration data sets, we show their time series around JD2453283 and JD2456195 as done for Fig. \ref{fig:2epochs}. We see strong correlation between RV and various proxies over times scales longer than one epoch. The Pearson correlation coefficients between RVs and proxies for the epochs around JD2453283 are -0.40, -0.36, 0.28, -0.02 and 0.02 for the $M_0$, $M_{c3}$, $M_{c4}$, $M_{c5}$ and cC3AP2-1 data set respectively, while the coefficients are 0.06, -0.73, 0.43, -0.50 and 0.32 for the other epochs. The rapid increase and decrease during these two epochs are clearly seen in other proxies. Thus the instrumental bias is probably the main reason why we find high eccentricity in Keplerian solutions even though the application of the noise model is able to significantly reduce such noise (see Fig. \ref{fig:2epochs}). However, there are also strong variations in moments within one observation night, which do not appear in the RVs. This short-term variation may be caused by the change of $\tau$ Ceti's altitude which modulates the intensity of the atmospheric absorption of the stellar light. Further studies are required to understand these variations before using the moments as noise proxies. In Fig. \ref{fig:moments_2epochs}, we observe that the c3AP2-1 data set decreases around JD2456195 as the RVs do. Since no RVs of the 172 stars were measured or published around epoch JD2453283, we assign the calibration data points at nearby epochs to them.

Although the moments correlate with RVs up to 0.5\,m/s, the inclusion of them in the model may have both positive and negative effects on the identification and quantification of signals because of the possibly incomplete subtraction of Keplerian variation from moments. Considering these problems, we have only used them to test the sensitivity of signals to noise models.

\section{Planetary candidates}\label{sec:candidates}
\subsection{Four planet system}\label{sec:four}
\begin{figure}
  \centering 
\includegraphics[scale=0.35]{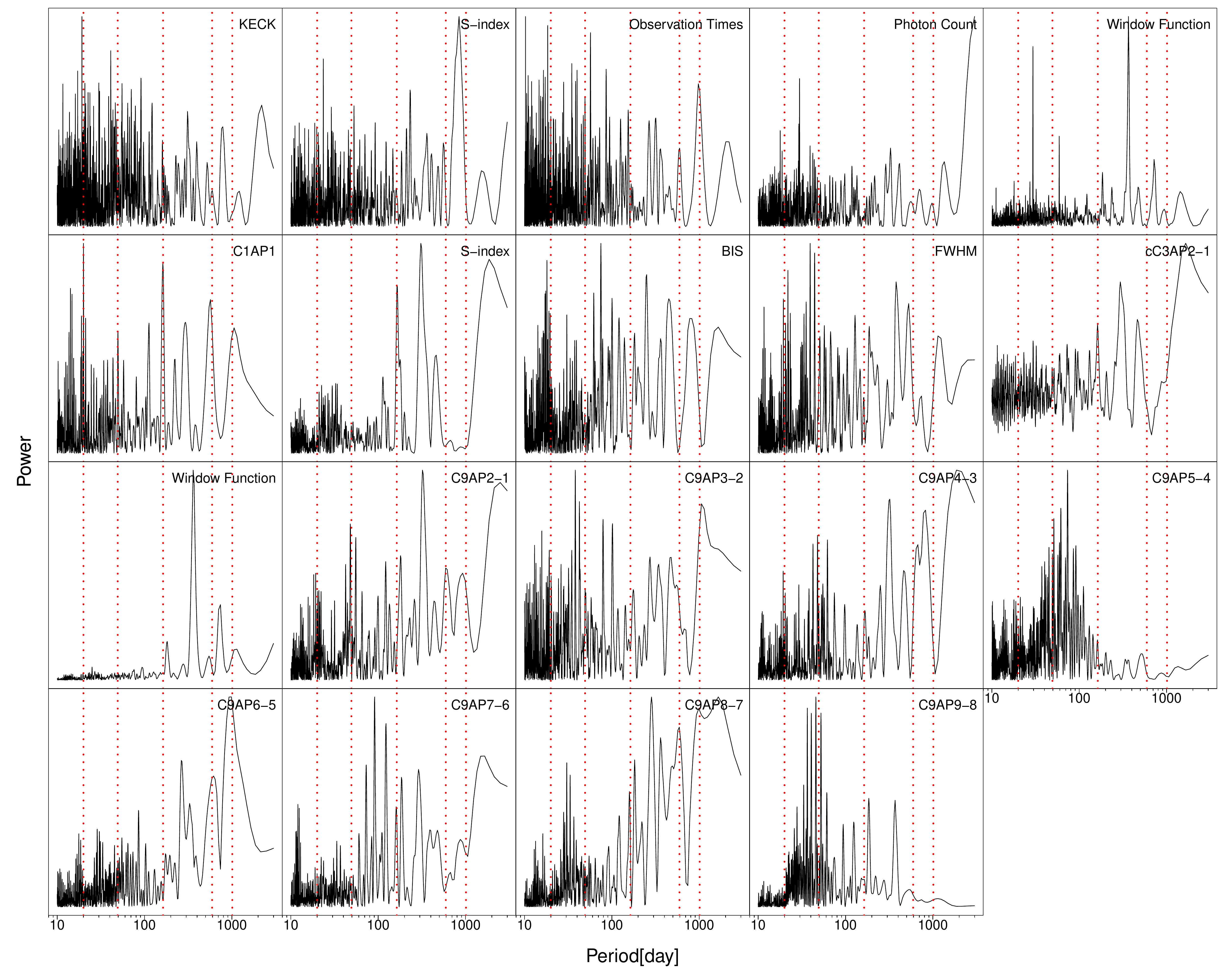}
\caption{The GLSTs of the data, activity indices of KECK (top panels), C1AP1 (middle and bottom panels) and differential RVs (bottom panels). The 600 and 1000\,d signals and the signals constrained in combination with the 600\,d signal for the C1AP1 data set (see the fifth column of Table \ref{tab:signals}) are shown by the red dotted lines. We have truncated the periodograms at a period of 10\,d to optimize visualization.}
\vspace*{0.1in}
\label{fig:activity}
\end{figure}

According to the analyses in section \ref{sec:signals}, we conclude that the signals with periods of 20, 49, 160 and 600 d are genuine Keplerian candidates. To further confirm them as Keplerian candidates, we check whether these signals could be caused by activity. We show the periodograms of the C1AP1 and KECK data sets, activity indices, differential RVs, and the signals in Fig. \ref{fig:activity}. Although there are some peaks around 600\,d and 1000\,d in the periodograms of observation times of the KECK data, 9AP6-5 and 9AP2-1, they are either minor peaks or only strong in one or two noise proxies. Notably the 1000\,d signal is rather strong in the periodogram of 9AP6-5 despite a considerable probability of random overlap between the signals in the data and the powers in noise proxies. The 160\,d signal is strong in the S-index but does not appear in other proxies. The periodograms both for C1AP1 and for KECK show strong peaks around 20\,d, strongly suggesting it as a genuine Keplerian signal. Other signals are not found to be strong in the KECK periodogram probably due to the large uncertainty and small sample of KECK measurements. These signals have mean semi-amplitudes below 0.5\,m/s \footnote{The mean semi-amplitudes of the Keplerian signals are equal to the semi-amplitudes determined in the circular solutions which are shown in the right columns of Table \ref{tab:signals}.}, which is beyond the detection ability of KECK.

Since the 600\,d signal has low eccentricity in the Keplerian solution for the C1AP1 data set, we show the parameters of the five signals identified in it in Table \ref{tab:signal_parameter}. The boundary of uncertainty intervals, $\theta_{1\%}$ and $\theta_{99\%}$, are determined at the cumulative marginalized posterior probability of 1\% and 99\%. Specifically, they are determined by $\int_{\theta_{min}}^{\theta_{1\%}}P(\theta|\mathcal{D})d\theta=1\%$ and $\int_{\theta_{min}}^{\theta_{99\%}}P(\theta|\mathcal{D})d\theta=99\%$, where $\theta_{min}$ is the minimum value of parameter $\theta$\footnote{Actually, the formula expressed here should be expressed in a discrete form because we calculate the intervals using MCMC samples.}. The uncertainties of parameters are probably larger than the reported values because the mean values of parameters slightly depend on the choice of data set as shown in Table \ref{tab:signals}. In Table \ref{tab:signal_parameter}, we observe that all of the five planets have minimum masses less than 5\,$M_\Earth$. In particular, $\tau$ Ceti g and h have minimum masses comparable with the Earth. $\tau$ Ceti h, e, and f have considerable eccentricities. The causes could be the instrumental noise we have investigated in section \ref{sec:signals} and the bias in the estimation of eccentricities since eccentricity is positive definite \citep{zakamska11}. The planetary masses would be double the minimum masses if the best-fitting inclination for the debris disk of $\tau$ Ceti (around 30$^\circ$) is equal to the inclination of the planetary system \citep{lawler14}. Our detection of these small planets at such long orbits demonstrates the potential power of the RV technique in finding Earth analogs. The RV variations caused by these planets have mean semi-amplitude as low as 0.3\,m/s which is close to the 0.1\,m/s limit required for detecting Earth analogs around solar analogs.

If we regard the measurements in a 15\,min time bin as an independent observations (e.g. \citealt{mayor03} and \citealt{otoole07}), there would be 662 observations for the C1AP1 data set. The $K/N=K/RV_{\rm rms}\times \sqrt{N_{\rm obs}}$ ratio for signals with $K=0.3$\,m/s is 7.3, where $RV_{\rm rms}$ is the standard deviation of the RVs after removing the best fitted trend and the correlation with noise proxies. This is close to the detection threshold of 7.5 based on the RV-challenge results \citep{dumusque16b}. Considering that our team has detected signals with $K/N=5$ without announcing false positives in the RV challenge, our detection of small signals is reliable and is consistent with previous analyses. 

\begin{table}
  \centering
\caption{The MAP estimation of the parameters for four signals detected in the C1AP1 data set. The uncertainties of parameters are denoted by the lower and upper bounds of the intervals, which are determined at the 1\% and 99\% quantiles of the posterior density. To estimate the semi-major axis $a$ and the minimum mass of planet $m \sin{i}$, we set the mass of $\tau$ Ceti to be $0.783\pm0.012M_\odot$ \citep{teixeira09}. To be consistent with the names used by MT13, the new signals at periods of 20 and 49\,d are named $\tau$ Ceti g and h. }
\label{tab:signal_parameter}
\footnotesize{
  \begin{tabular}  {c*{4}{c}}
\hline 
Parameters&    $\tau$ Ceti g& $\tau$ Ceti h& $\tau$ Ceti e& $\tau$ Ceti f\\\hline
    $P$\,(d)&20.00 [19.99, 20.02]&49.41 [49.31, 49.49]&162.87 [162.41, 163.95]&636.13 [588.44, 647.83]\\
    $K$\,(m/s)&0.49 [0.38, 0.56]&0.39 [0.33, 0.54]&0.55 [0.46, 0.68]&0.35 [0.23, 0.45]\\
    $e$&0.06 [0.00, 0.19]&0.23 [0.08, 0.39]&0.18 [0.04, 0.36]&0.16 [0.00, 0.23]\\
    $\omega$\,(rad)&6.90 [5.42, 7.53]&0.13 [-0.71, 0.78]&0.39 [-0.41, 1.68]&2.09 [0.86, 2.81]\\
    $M_0$\,(rad)&7.04 [6.67, 8.65]&-1.27 [-1.84, -0.52]&6.21 [5.27, 7.37]&-0.68 [-1.32, 0.30]\\
    $m\sin{i}$\,($M_\Earth$)&1.75 [1.35, 2.00]&1.83 [1.57, 2.51]&3.93 [3.29, 4.76]&3.93 [2.56, 4.98]\\
    $a$\,(au)&0.133 [0.131, 0.134]&0.243 [0.240, 0.246]&0.538 [0.532, 0.544]&1.334 [1.290, 1.351]\\
\hline
  \end{tabular}
  }
\end{table}

\begin{figure}
\centering
\includegraphics[scale=0.6]{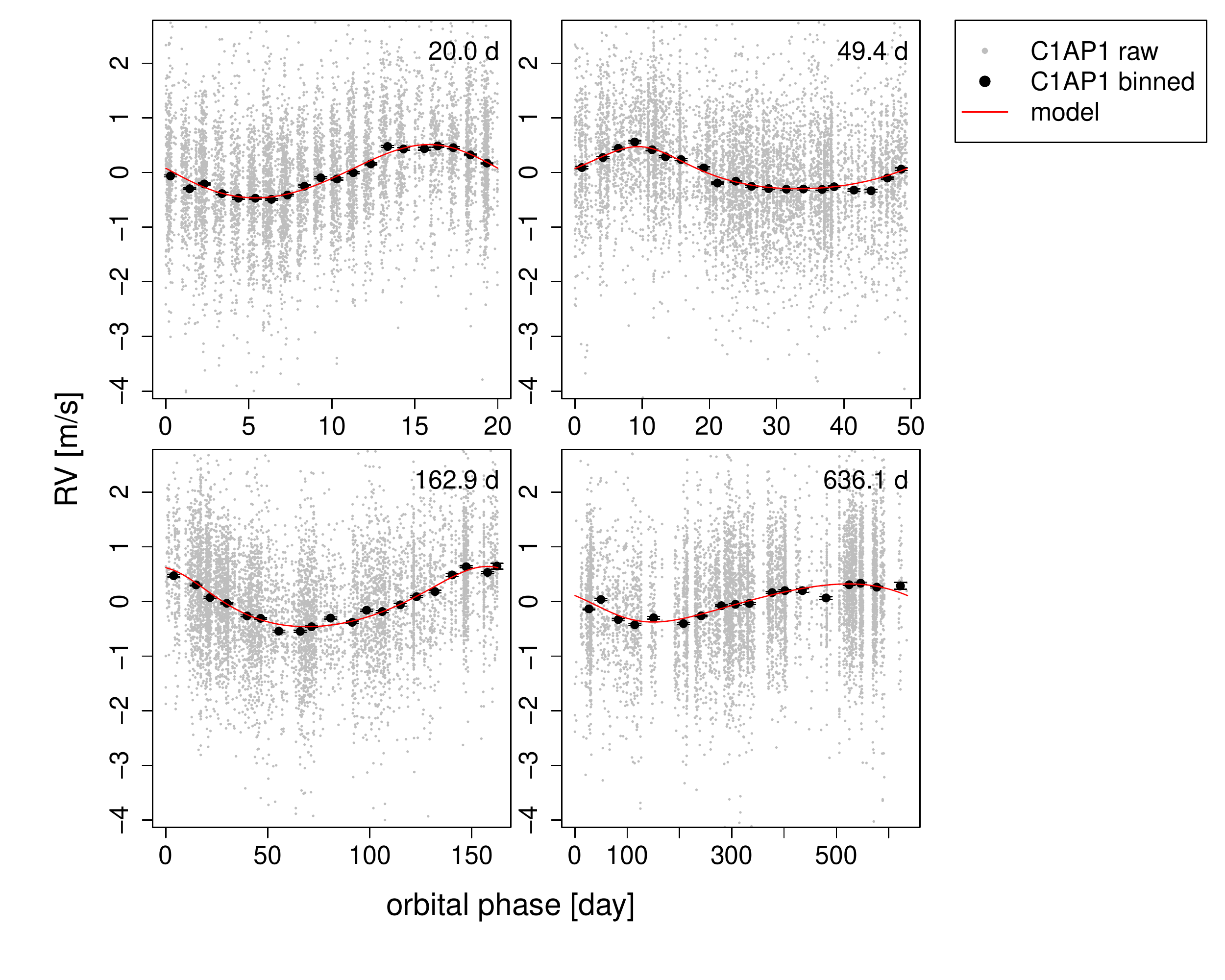}
  \caption{The phase-folded data and the best model prediction for each signal for the C1AP1 data set. The data and the binned version are shown as small and big dots. The MAP estimation of the four signals are based on 8 million posterior samples drawn by a cold chain. }
  \label{fig:phase_fold}
\end{figure}

\begin{figure}\centering 
\includegraphics[scale=0.6]{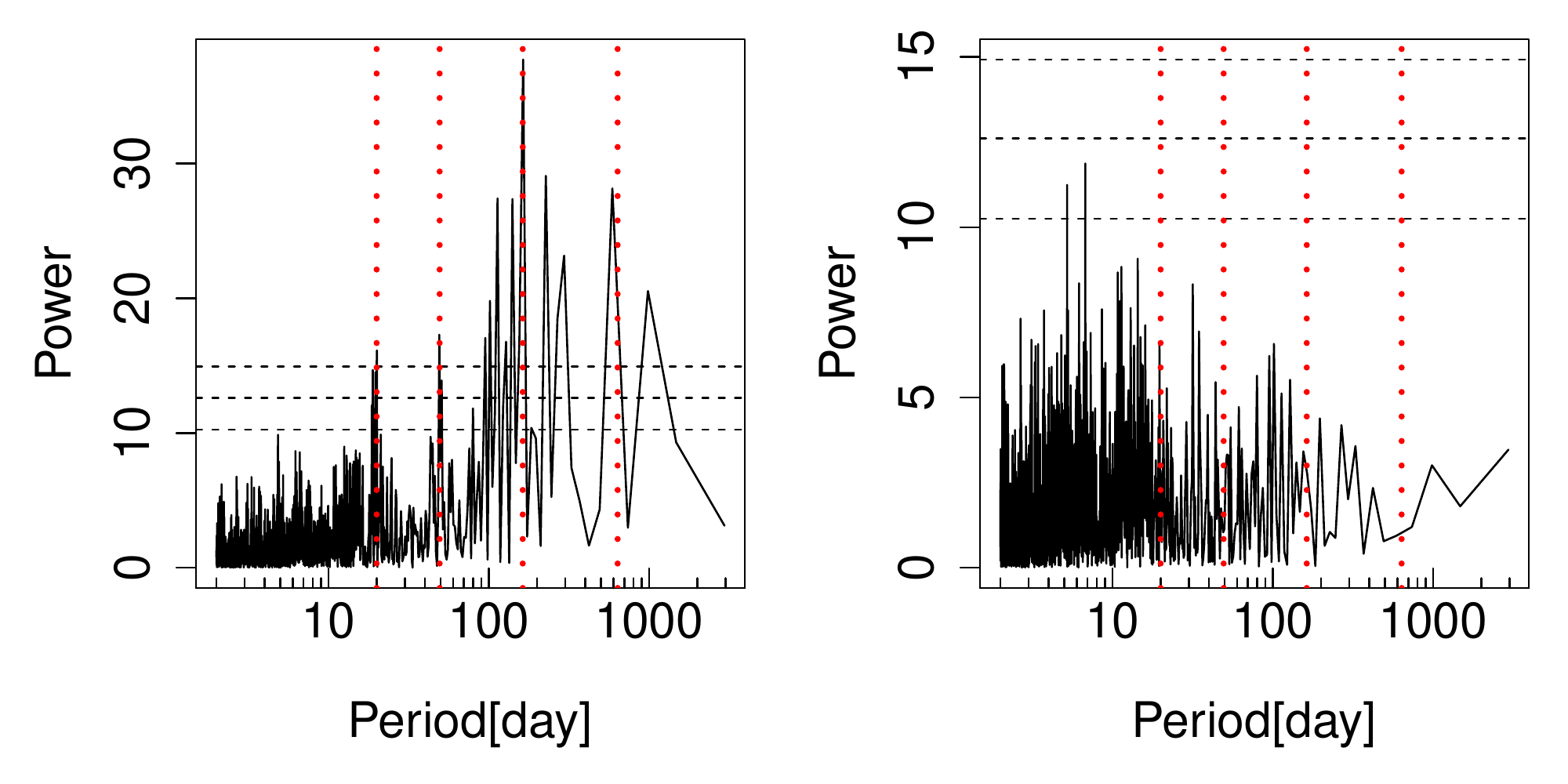}
  \caption{The GLSTs for the C1AP1 data set after subtracting the best-fitting 0P+MA(4)+8D (left) and 5P+MA(4)+8D (right) model. The signals are denoted by the red dotted lines. Since the correlated noise is subtracted, the FAPs of 0.1, 0.01 and 0.001 are shown by dashed lines as a metric to measure the significance of signals. }
\label{fig:all_residual}
\end{figure}

We also show the phase-folded data and model prediction for all signals in Fig. \ref{fig:phase_fold}. We see that the data strongly support the existence of four signals. 
To see whether there are extra signals in the data, we subtract the best-fitting model (the 600\,d Keplerian solution shown in Table \ref{tab:signals}) from the C1AP1 data, and show the GLST of the residual in Fig. \ref{fig:all_residual}. We do not see any strong signals, supporting the fact that the six-planet model is not favored by the data. 

Apart from the HARPS data set, we also combine the RVs measured by the Anglo Australian Telescope (AAT; the data is available in MT13) and KECK with HARPS. We find results consistent with the five planet solution. Thus the weak wobbles found in HARPS data are consistent with other data despite not being identified independently due to relatively lower RV precision of other instruments. 

\subsection{Comparison with previous studies}\label{sec:previous}
The planetary candidates we have identified partially overlap with the ones found by MT13. We find two new planetary candidates with periods around 20 and 49\,d, but fail to confirm the signals around 14 and 35\,d. Although there is evidence for the existence of the signal at around 92\,d, we cannot confirm it as a Keplerian candidate because it cannot be consistently identified in all data sets and solutions. The signal around 14\,d becomes weak when we subtract the 20\,d signal from the data. But the opposite is not true, suggesting a non-Keplerian origin of the 14\,d signal. Nevertheless, the 14\,d signal does exist in some Keplerian and circular solutions after accounting for the 20\,d signal. In addition, by evenly dividing the data into 3 chunks, we find that the 14\,d is significant in the first chunk while the 20\,d is significant in the second and third chunks. Since MT13 only analyzed the first two chunks, it is expected that they find the 14\,d signal to be favored. However, as we have mentioned in section \ref{sec:curse}, the first chunk is problematic because of its high jitter level and the abnormal variation of FWHM before JD24503280. Moreover, this chunk contains high-cadence asteroseismology data which is influenced by guiding errors. Considering these factors, the 14\,d signal is probably an activity-induced signal rather than an alias of 20\,d. 

The 35\,d signal reported by MT13 or the 32\,d signal identified in this work is very weak in the C1AP1 data set (see Fig. \ref{fig:primary}). Moreover, the 35\,d signal identified by MT13 is close to the rotation period of 34\,d determined by \cite{baliunas96}, although this rotation period is not significant in the periodograms for activity indices and for differential RVs (see Fig. \ref{fig:activity}). We also find that the alias of the 600\,d signal, located at around 1000\,d, fits the data as well as the 600\,d signal if high eccentricity is allowed (see Table \ref{tab:BFs}).

The semi-amplitudes of signals identified in this work are from 0.4 to 0.5\,m/s. These values are lower than the values reported by MT13, which are from 0.58 to 0.75\,m/s. In the analysis of MT13, the activity-induced variation is probably misinterpreted as signals because of insufficient noise modeling. Moreover, the periods of signals in this work are different from the values reported by MT13. This difference together with the other differences are probably caused by the following factors. 

First, the HARPS RV sample used in this work is double of the size of the sample (4398 RV points) used by MT13. Although MT13 have also used the KECK and AAT samples, both of them have larger uncertainties and have less than 1000 measurements. The updated HARPS data set allows us to find weaker signals, and quantify signals more accurately. In addition, the early HARPS data are found to be more noisy than recent ones according to our analyses. This could also be the reason why MT13 have identified somewhat different signals. 

Second, we use the differential RVs to remove wavelength dependent noise in the data. Although the aperture data sets are not available for the KECK data, the signal is mainly constrained by the HARPS data (see Fig. \ref{fig:phase_fold}) where the noise is properly modeled by differential RVs (see Fig. \ref{fig:residual}). In addition, we test different noise models and select the best one for each data set to avoid false negatives and positives. 

Third, we estimate the BF using the BIC which penalize complex models more strongly than the Akaike information criterion (AIC) and the truncated posterior mixture (TPM) used in MT13. According to the comparison of various BF estimators by \cite{feng16}, the BIC is more conservative in estimating the BF than the AIC and TPM estimators because it assumes that all parameters are equally free. However, this might not be appropriate due to the nonlinear combination of Keplerian parameters, leading to the unequal ability of parameters to improve the fitting. For example, the likelihood is not as sensitive to the eccentricity as to the period. This can be seen in Table \ref{tab:signal_parameter} where the relative uncertainty of eccentricity is much larger than that of the period. As there is no universal method to calculate the BF, we regard the BIC as a practical and conservative diagnostic tool to confirm signals. 

\subsection{Dynamical stability and habitability analysis}\label{sec:stability}
\begin{figure}\centering 
\includegraphics[scale=0.6]{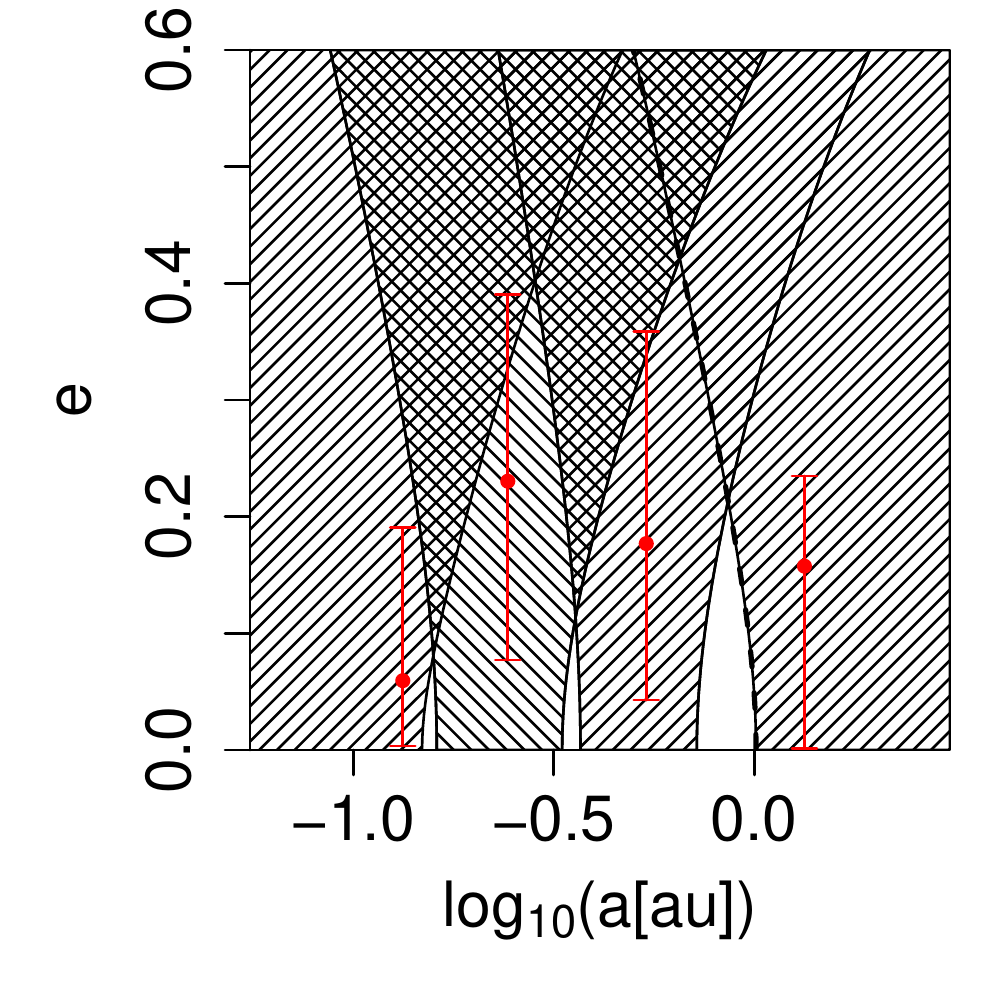}
  \caption{The Lagrange stability of planets around $\tau$ Ceti. The shaded region represents the unstable orbits around a planet. The signals around 20, 49, 160 and 600\,d are quantified simultaneously in the 5-planet solution for the C1AP1 data set, and are denoted by red dots with error bars (measured at the 1\% and 99\% quantiles of posterior density). The shaded regions denote the parameter space of unstable orbits. }
\label{fig:lag}
\end{figure}

We now assume that the detected signals correspond to planets orbiting $\tau$ Ceti. We analyze the dynamical stability of the planetary candidates using the threshold of Lagrange stability introduced by \cite{barnes06}. We calculate the instability boundary in the phase space of eccentricity and semi-major axis using the MAP estimation of the parameters of candidate planets, and show the results in Fig. \ref{fig:lag}.
Although the orbits of candidates are eccentric, the system is dynamically stable. As we have shown in section \ref{sec:instrument}, the high eccentricity is probably caused by instrumental noise. The eccentricity of the planetary candidates might be over-estimated because eccentricity is positive definite \citep{zakamska11}. Considering the overestimation of eccentricity, the stable region (or the gap between shaded regions) could be larger than that in the figure. On the other hand, the stable region would shrink if the true masses of these candidates are much larger than their minimum masses. But the area of the shaded regions are more sensitive to eccentricity than to the mass according to Eqn. (2) in \cite{barnes06}. Therefore we consider the Lagrange stability shown in Fig. \ref{fig:lag} as a conservative illustration of the dynamical stability of the planetary system. 

The four planet system is dynamically packed, which is probably due to the effect of planet-planet scattering \citep{raymond09}. Our result is consistent with an inner edge as low as 1.6\,AU determined by ALMA observations \citep{macgregor16}. The dynamical simulations of the planets reported by MT13 show that the previously reported five planet system is stable over long time scales \citep{lawler14}. This indicates that the four planet system reported in this work may also be stable for a long time since both of the two systems are tightly packed and the outer planets have similar semi-major axes. 

We analyze the habitability of the planets by calculating the habitable zone according to the method introduced by \cite{kopparapu14}. We adopt the effective temperature and luminosity of the $\tau$ Ceti measured by \cite{santos04} and \cite{pijpers03}, which are 5,344\,K and 0.52\,$L_\odot$, respectively. We find, for a one Earth mass planet, the conservative habitable zone of $\tau$ Ceti is from 0.70 to 1.26\,au while the habitable zone for a five Earth mass is from 0.68 to 1.26\,AU. Thus none of the reported planets reside in this conservative zone. If one instead takes the recent Venus and early Mars limits of the habitable zone as 0.55 and 1.32 AU then $\tau$ Ceti e and f are close to the inner and upper boundary of the optimistic habitable zone. However, $\tau$ Ceti e and f are probably not dynamically habitable. The ALMA observations of $\tau$ Ceti suggests the existence of a broad debris disk \citep{macgregor16}. The large amount of asteroids and comets in the disk may be strongly perturbed by planets, leading to an impact rate ten times higher than that on the Earth \citep{greaves04}. On the other hand, the impact rate on these planets may depend not only on the mass of scattered disc but also on the structure of the planetary system and the interstellar environment \citep{feng14}. Thus comprehensive studies on the dynamics of the $\tau$ Ceti system is important for understanding its habitability. 

\section{Discussions and Conclusions}\label{sec:conclusion}

We analyze the RV data of $\tau$ Ceti in the Bayesian framework. We find strong dependence of the RV noise on wavelengths or orders. This colorful noise cannot be removed by averaging all spectral orders, and would cause inconsistent results if not removed. To model this noise, we introduce a wavelength dependent noise model by linearly combining moving average models with differential RVs. We apply this model to aperture data sets, and find that the differential RVs efficiently remove wavelength dependent noise, reducing the false positive and negative rate. We also find that binning RVs over time would change both the signals and noise in the data due to the lack of an appropriate weighting function. Therefore we propose a combination of moving average models and differential RVs to model the time and wavelength dependent noise in unbinned RV datasets. 

Our findings challenge the traditional methods which only account for wavelength independent noise. It means that the planetary candidates with semi-amplitude of around 1\,m/s found in previous studies may require further confirmation by applying differential RVs to noise modeling. New data reduction algorithms are needed to extract aperture data sets from Doppler measurements for various instruments. Another application of differential RVs is concerned with the diagnostics of stellar activity. Aperture data sets provide abundant information for the study of stellar physics. The spectral powers of differential RVs are wavelength dependent (as shown in Fig. \ref{fig:phase_fold}), suggesting the colorfulness of stellar activity. Further investigations are necessary to confirm differential RVs as signatures of the spectral and photometric variations of stellar surfaces. On the other hand, our work focuses only on the analysis of the large RV data for $\tau$ Ceti. Further studies on the connection between differential RVs and stellar activity for various types of stars and instruments are needed to extend the wavelength dependent noise model proposed in this work. 

We apply the noise model to various RV data sets and get Keplerian and circular solutions. Both solutions consistently identify the signals around 600, 160, 20 and 49\,d for all data sets. For the Keplerian solutions, the strongest signal has a period of either around 600\,d or around 1000\,d, which are aliases of each other. The 600\,d signal is probably a genuine signal because it is favored over 1000\,d for the circular solutions. Moreover, the eccentricity of the former is low for some data sets while the latter is highly eccentric for all data sets. Nevertheless, significantly eccentric signals exist in all Keplerian solutions. The relevant results and data are available at \url{http://star-www.herts.ac.uk/~ffeng/HD10700_supplementary/tau_ceti_results/}.

By dividing the data into chunks, we see the sensitivity of data to signals varying with time. But all signals can be found if we fit the same Keplerian component and independent noise components to data chunks. Moreover, we have also studied the instrumental noise in HARPS, and find that the HARPS data is biased at the level of at least 0.2\,m/s. The central moments of the spectral line profile are also found to be strongly correlated to the data at the level of at least 0.5\,m/s. Future studies are required to confirm whether or not the correlation is caused by the Keplerian variation left in the central moments due to the dependence of RVs to spectral flux. The signals we have identified are robust to the inclusion of central moments in the noise model.

The planetary candidates found in this work are partially overlap with those detected by MT13. With more HARPS measurements and wavelength dependent noise models, we improve the estimation of the 600 and 160 signals identified by MT13. We also find two new planets together with the 1000\,d alias of the 600\,d signal but fail to confirm the other two planetary candidates proposed by MT13. Since the signal at a period of 92\,d cannot be consistently identified in all data sets, we don't confirm it as a planetary candidate. These differences suggest that we should be very cautious about the interpretation of sub-m/s signals detected in large and high cadence data sets. For example, the signals identified around $\alpha$ Centauri B \citep{dumusque12}  are found to arise from the window function \citep{rajpaul15}. 

Our analysis of the dynamical stability of the four planet system suggests that $\tau$ Ceti h, e, and f should have eccentricities lower than their values determined in this work if they are genuine planets. Although $\tau$ Ceti e and f are located close to the boundaries of the optimistic habitable zone, their habitability might be strongly reduced by the bombardments of objects from the massive scattered disc. The determination of the inner boundary of this scattered disc may help to confirm the existence of the 600\,d candidate. If the scattered disc and the planetary system are coplanar, the true masses of these planetary candidates are double the minimum masses reported here \citep{lawler14,macgregor16}.

Our detection of Keplerian signals as low as 0.4\,m/s demonstrates the unique role of the RV technique in detecting Earth analogs around Sun-like stars. Unlike the transit technique, the RV method does not require the occurrence of transits, and can be useful to detect planets around all bright stars. With the development of high precision spectrometers, advanced statistical methods and noise models, it is feasible to find Earth analogs in the coming decade. Since the weak wobbles caused by Earth analogs are comparable to the measurement uncertainties and jitter levels, it is crucial to build appropriate noise models to avoid both false positives and negatives. Our work on modeling wavelength dependent noise is intended as a step toward a noise model framework where physics-motivated models and stochastic processes are properly combined to remove RV noise. 

\appendix 
\section{Correlation between noise proxies}\label{sec:correlation}
\begin{figure}
  \centering
   \hspace*{-15mm}
  \includegraphics[scale=0.5]{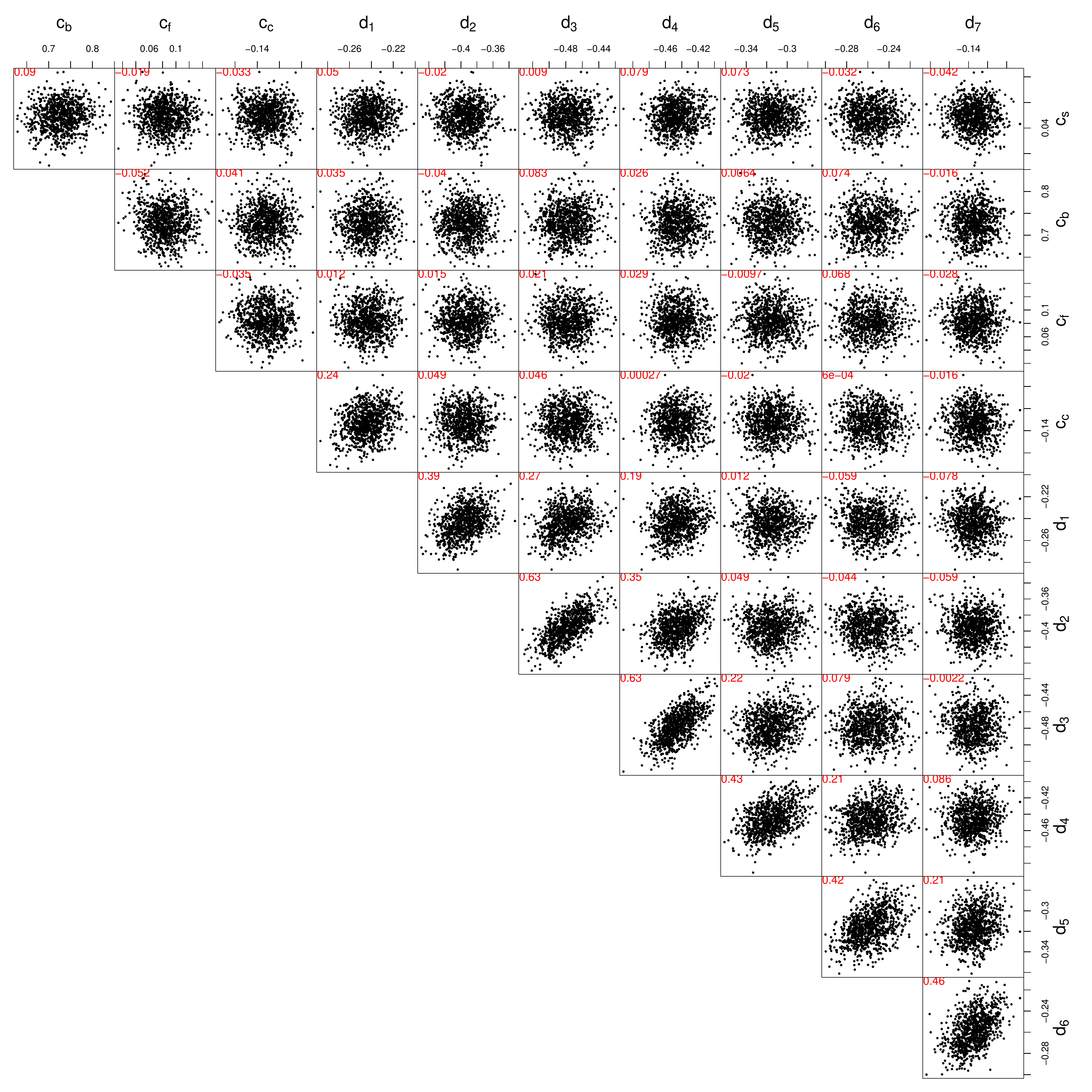}
  \caption{The correlation between noise model parameters based on 1000 samples randomly drawn from 4 million posterior samples of the 4P+MA(4)+8D model for the C1AP1 data set. In this Keplerian solution, the signals have periods of 49.5, 91.7, 20 and 597\,d. The Pearson correlation coefficient is shown in the top left corner for each panel. The parameters of $c_s$, $c_b$, $c_f$ and $c_c$ are the linear coefficients of S-index, BIS, FWHM and cC3AP2-1. The other parameters are linear coefficients of differential RVs derived from 9AP aperture data sets. All parameters are in units of m/s.}
  \label{fig:pair}
\end{figure}

We test the correlation between the linear coefficients of noise proxies during the fitting. We randomly select 1000 samples from the MCMC chain for the 4P+MA(4)+8D Keplerian solution (with the 600\,d signal identified) for the C1AP1 data set and show the correlations between the linear coefficients of various noise proxies in Fig. \ref{fig:pair}. Most noise proxies are independent of one another. There are weak correlations between differential RVs derived from nearby spectral orders because of much common noise shared by them. Nevertheless, such a correlation may not cause the so-called ``collinearity'' problem, considering that the MCMC chains show no evidence for non-convergence. Moreover, the application of noise proxies in fitting is seen in the analysis of Kepler light curves (e.g. \citealt{foreman-mackey15}).

\section{Wavelength-dependence of model parameters}\label{sec:weight}
\begin{figure}
  \centering
   \hspace*{-10mm}
  \includegraphics[scale=0.5]{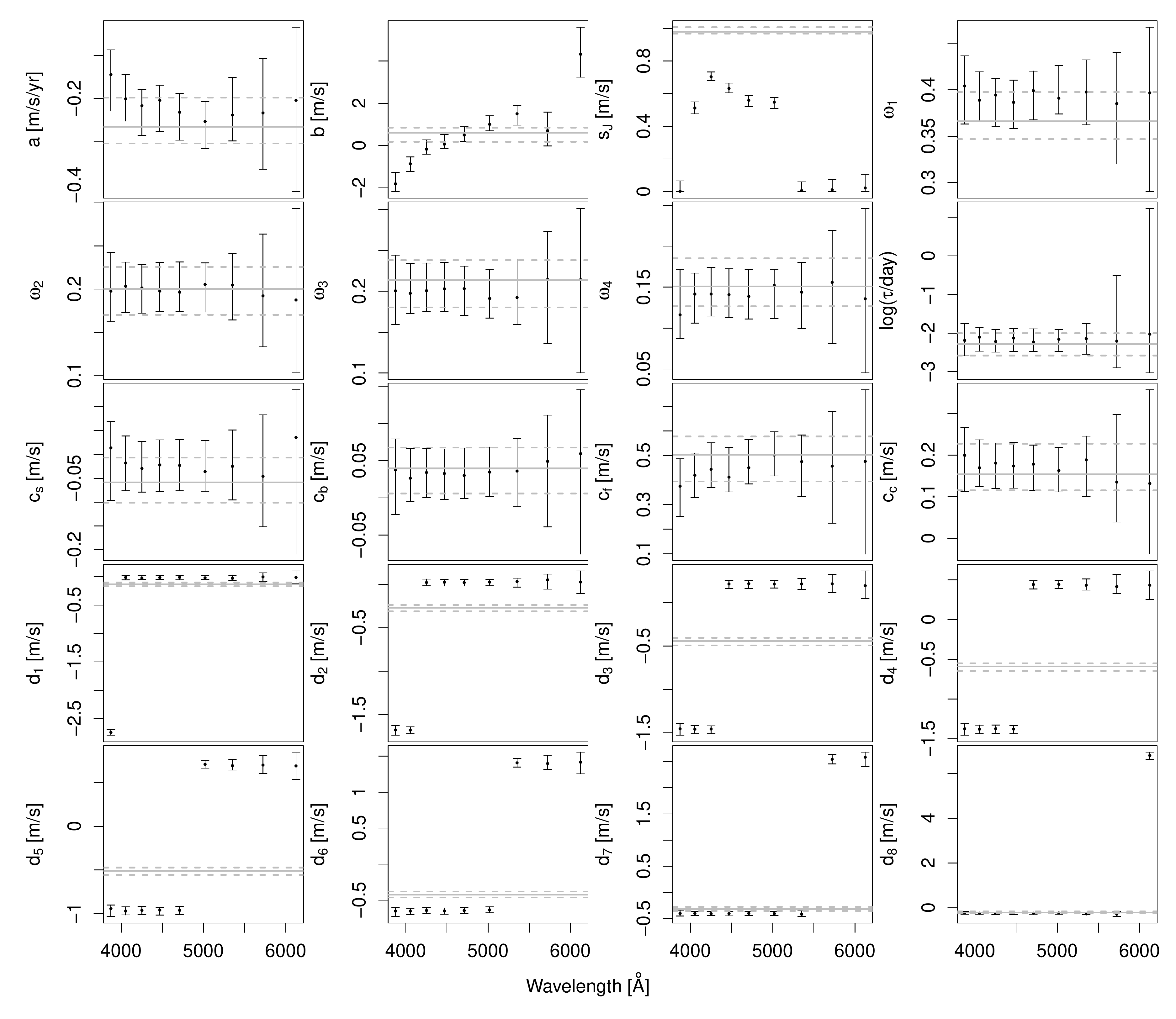}
  \caption{The MAP estimation of parameters varying with wavelength for the 9AP data sets. The error bars are determined by the 1\% and 99\% quantiles of the posterior densities. The MAP estimation of parameters for the 1AP1 data set and corresponding uncertainties are denoted by solid and dashed lines, respectively. The variation of $d_i$ ($i\in\{1, ..., 8\}$) with wavelength is step-like because two differential RVs with close wavelength ranges are derived from the same aperture data set but with opposite signs. } 
  \label{fig:wave}
\end{figure}

To explore the wavelength dependence of noise model parameters, we apply the 0P+MA(4)+8D model to 9AP aperture data sets. We show the dependence of the MAP estimations of parameters and uncertainties on the mean wavelength of aperture data sets in Fig. \ref{fig:wave}. We see strong wavelength dependence of jitter $s_J$ and reference velocity $b$. Thus a simple average of all spectral orders would introduce wavelength-dependent noise without a proper weighting function. We also see weak wavelength dependence of the trend $a$ and FWHM $c_f$. Notably we observe strong wavelength dependence of differential RVs. For a given aperture data set, the coefficients for the differential RVs with shorter and longer wavelengths always have opposite signs because they are adjusted to fit the aperture data set. For example, the coefficients for 9AP2-1, 9AP3-2, 9AP4-3, 9AP5-4, 9AP6-5 are positive while the other differential RVs are negative in the analysis of 9AP6. Hence differential RVs have ``re-weighted'' aperture data sets such that the color difference between aperture data sets disappears. This is also the reason why the periodogram difference between aperture data sets disappear if including differential RVs in the model (see Fig. \ref{fig:residual}).
\label{app}

\section*{Acknowledgements}
FF, MT and HJ are supported by the Leverhulme Trust (RPG-2014-281) and the Science and Technology Facilities Council (ST/M001008/1). We used the ESO Science Archive Facility to collect radial velocity data sets.  The authors gratefully acknowledge the HARPS-ESO team’s continuous improvement of the instrument, data and data processing that made this work possible. Finally, the authors would like to thank the anonymous referees for their valuable comments that enabled considerable improvements of the manuscript. 

\bibliographystyle{aasjournal}
\bibliography{nm}  
\end{document}